\documentclass[pra, aps, superscriptaddress, twocolumn, longbibliography]{revtex4-2}

\usepackage{array,multirow,graphicx,color,amsmath,amsfonts,enumerate,amsthm,amssymb,mathtools,enumitem,thmtools,hyperref,  mathdots,enumitem,centernot,bm,soul,bbm,tikz,pgfplots}
\usepackage{enumitem}
\newcolumntype{P}[1]{>{\centering\arraybackslash}p{#1}}
\usepackage{multirow}
\usepackage[capitalise, noabbrev]{cleveref}
\usetikzlibrary{arrows}
\pgfplotsset{compat=1.14}
\hypersetup{colorlinks=true,linkcolor=blue,citecolor=blue,urlcolor=blue}

\usepackage{braket}

\theoremstyle{plain}

\theoremstyle{definition}

\newcommand{\ect}{E_{C, t}}
\newcommand{\ecf}{E_{C, f}}
\newcommand{\elf}{E_{L, f}}
\newcommand{\ejf}{E_{J, f}}
\newcommand{\ejt}{E_{J, t}}
\newcommand{\omt}{\omega_{t}}
\newcommand{\oms}{\omega_{s}}
\newcommand{\anharm}{\delta_{t}}
\newcommand{\omf}[1]{\omega_{f, #1}}

\newcommand{\qf}{q_{f}}
\newcommand{\phif}{\phi_{f}}
\newcommand{\qt}{q_{t}}
\newcommand{\qzpf}{q_{\mathrm{zpf}}}
\newcommand{\qfel}[2]{q_{f, #1 #2}}

\newcommand{\omd}{\omega_{d}}
\newcommand{\epsd}{\varepsilon_{d}}
\newcommand{\phased}{\theta_d}
\newcommand{\jc}{J_C}
\newcommand{\ketbra}[2]{\ket{#1}\bra{#2}}
\newcommand{\ketbrat}[2]{\ket{#1}\bra{#2}_t}
\newcommand{\ketbraf}[2]{\ket{#1}\bra{#2}_f}
\newcommand{\sigmat}[2]{\sigma_{#1 #2}^t}
\newcommand{\sigmaf}[2]{\sigma_{#1 #2}^f}

\begin{document}

\preprint{APS/123-QED}

\title{Microwave-activated gates between a fluxonium and a transmon qubit}

\author{A. Ciani}
 \affiliation{Institute for Quantum Computing Analytics (PGI-12), Forschungszentrum J\"ulich, 52425 J\"ulich, Germany}
 \affiliation{QuTech, Delft University of Technology, P.O. Box 5046, 2600 GA Delft, The Netherlands}
 \author{B. M. Varbanov}
\affiliation{QuTech, Delft University of Technology, P.O. Box 5046, 2600 GA Delft, The Netherlands}
\author{N. Jolly}
\affiliation{QuTech, Delft University of Technology, P.O. Box 5046, 2600 GA Delft, The Netherlands}
\affiliation{Mines ParisTech, PSL Research University, F-75006 Paris, France}
\author{C. K. Andersen}
\affiliation{QuTech, Delft University of Technology, P.O. Box 5046, 2600 GA Delft, The Netherlands}
\affiliation{Kavli Institute for Nanoscience, Delft University of Technology, 2600 GA Delft, The Netherlands}
\author{B. M. Terhal}
\affiliation{QuTech, Delft University of Technology, P.O. Box 5046, 2600 GA Delft, The Netherlands}
\affiliation{JARA, Institute for Quantum Information, Forschungszentrum J\"ulich, 52428 J\"ulich, Germany}
\affiliation{EEMCS, Delft University of Technology, Mekelweg 4, 2628 CD Delft, The Netherlands}

\date{\today}

\begin{abstract}
We propose and analyze two types of microwave-activated gates between a fluxonium and a transmon qubit, namely a cross-resonance (CR) and a CPHASE gate. The large frequency difference between a transmon and a fluxonium makes the realization of a two-qubit gate challenging. For a medium-frequency fluxonium qubit, the transmon-fluxonium system allows for a cross-resonance effect mediated by the higher levels of the fluxonium over a wide range of transmon frequencies. This allows one to realize the cross-resonance gate by driving the fluxonium at the transmon frequency, mitigating typical problems of the cross-resonance gate in transmon-transmon chips related to frequency targeting and residual $ZZ$ coupling. However, when the fundamental frequency of the fluxonium enters the low-frequency regime below $100 \, \mathrm{MHz}$, the cross-resonance effect decreases leading to long gate times. For this range of parameters, a fast microwave CPHASE gate can be implemented using the higher levels of the fluxonium. In both cases, we perform numerical simulations of the gate showing that a gate fidelity above $99\%$ can be obtained with gate times between $100$ and $300 \, \mathrm{ns}$. Next to a detailed gate analysis, we perform a study of chip yield for a surface code lattice of fluxonia and transmons interacting via the proposed cross-resonance gate. We find a much better yield as compared to a transmon-only architecture with the cross-resonance gate as native two-qubit gate.
\end{abstract}

\maketitle


\section{Introduction}\label{sec::intro}
The transmon qubit \cite{koch2007, schreier2008} is the most succesful superconducting qubit to date with superconducting chips with around a hundred qubits currently being realized \cite{wu2021, Arute2019, ibmq}. The success of the transmon is due to its resilience to charge noise, the relative simplicity of the circuit and its fabrication, the straightforward control and readout using microwave pulses, and the possibility to couple transmons either via direct capacitances \cite{Barends2014} or via bus resonators \cite{Majer2007, DiCarlo2009}. Coherence times between $10$ and $100 \, \mu s$ are routinely reported in two-dimensional transmon chips \cite{Andersen2020, Marques2022, ibmq}, and even longer $T_1$ times have been obtained by using different superconducting materials \cite{Place2021, wang2022}. High-fidelity two-qubit gates have been successfully demonstrated for transmon architectures using several different schemes that rely on either flux pulses \cite{DiCarlo2009, rol2019, negirneac2021, lacroix2020, barends2019, caldwell2018}, microwave drives \cite{sheldon2016, mitchell2021, krinner2020a, wei2021, kandala2021} or tunable couplers \cite{geller2015, yan2018, foxen2020, li2020, roth2017, mckay2016, collodo2020, sung2021}. 

Despite its success, the fact that transmons are essentially slightly anharmonic oscillators is a limiting factor in transmon architectures. Apart from the problem of leakage out of the computational subspace \cite{Varbanov2020}, the small anharmonicity of the transmon implies that the transmons must be separated in frequency by at most their anharmonicity to enable fast entangling gates. As observed in Ref.~\cite{magesan2020}, this is intuitively due to the fact that when the transmons are far away from each other in frequency, they behave as uncoupled harmonic oscillators. Notice however that in this frequency range also the unwanted, spurious $ZZ$ coupling is relatively large and this limits the performances of the gates \cite{sheldon2016, magesan2020, malekakhlagh2020}. These problems affect fixed, non-tunable coupling architectures such as those based on the cross-resonance (CR) gate \cite{rigetti2010, sheldon2016, magesan2020}, giving rise to the problem of frequency collisions \cite{Hertzberg2021}. The consequence is a low chip yield when qubit connectivity is as required for the surface code \cite{versluis2017}, prompting research into optimising the choice of qubit frequencies~\cite{morvan2021}, improving the accuracy with which qubit frequencies are targeted via laser annealing~\cite{Muthusubramanian2019APS, Hertzberg2021, Zhang2022} or pursuing alternative heavy-hexagonal codes that require a lower qubit connectivity~\cite{chamberland2020, Hertzberg2021}. Another issue in transmon chips is the problem of $ZZ$ crosstalk \cite{krinner2020b} for which some solutions have been discussed \cite{wei2021, mitchell2021, mundada2019, ku2020, sung2021, zhao2021, kandala2021}.

The fluxonium qubit \cite{Manucharyan113, ManucharyanPhd} is a suitable candidate to go beyond the limitations of a transmon-only architecture. The circuit of the fluxonium is similar to that of the transmon in being composed of a capacitance and a Josephson junction in parallel, but it also features an additional large, shunting inductance. The fluxonium is operated in the regime where the characteristic impedance of the parallel $LC$-circuit $Z_{LC} = \sqrt{L/C}$ is larger than, say, a few $ \mathrm{k\Omega}$s. To achieve this regime, the high impedance can be realized effectively as an array of hundreds of Josephson junctions \cite{ManucharyanPhd, masluk2012} or using a material such as granular aluminum \cite{Grunhaupt2019} or niobium-titanium-nitrate \cite{hazard2019, pitavidal2020}. 
The inductive shunt provides an intrinsic protection against charge noise, without the need of a large capacitance as in transmon qubits. Crucially, this breaks the tradeoff between anharmonicity and charge noise sensitivity that limits the transmon qubit. The large inductance also suppresses the sensitivity to flux noise in the loop formed with the Josephson junction (see Fig.~\ref{fig:tr_flx}a), and moreover, the fluxonium is usually operated in the double well configuration, where the qubit frequency is first-order insensitive to flux noise. In this configuration, the fluxonium qubit shows large coherence times \cite{nguyen2019} which have surpassed the millisecond barrier in 3D devices \cite{somoroff2021}. 

The enhanced protection of the fluxonium comes at the price of requiring a more involved scheme for the manipulation of its quantum state. The low fundamental frequency of the fluxonium (below $1 \mathrm{GHz}$) complicates the execution of single-qubit gates due to the lesser accuracy of the rotating wave approximation compared to the transmon case \cite{nguyen2022}. In addition, the reduced matrix elements of charge and flux operators, that control the strength of the coupling  between computational levels, needs to be compensated using higher drive power. At very low frequencies of around $10 \, \mathrm{MHz}$, new schemes have to be devised for state preparation (reset) and single-qubit gates \cite{zhang2021}. On the other hand, the measurement of a fluxonium qubit can have advantages as compared to a transmon qubit. The off-resonant fluxonium-resonator coupling can give rise to relatively-large dispersive shifts \cite{guanyu:dispersive,nguyen2022} which enable fast measurement. For granular aluminium based fluxonium qubits, highly-accurate quantum measurements using strong drive power -- populating the read-out cavity with a large number of photons -- have been reported \cite{gusenkova:flux-meas, takmakov:flux-meas}.

Despite this increased complexity, the higher coherence times reached by the fluxonium still pay off in terms of single-qubit gate fidelity \cite{somoroff2021}. Recently, several two-qubit gate schemes between fluxonia have been proposed and experimentally realized in two-qubit chips \cite{nesterov2018, ficheux2021, xiong2021, chen2021, nesterov2021, bao2021, dogan2022}. A whole architecture for fluxonium qubits has been analyzed in Ref.~\cite{nguyen2022} with two-qubit gates implemented using either the CR gate or the CPHASE gate induced by the differential AC-Stark shift effect \cite{mitchell2021}. 

In most superconducting qubit research, the focus is on coupling `same-type' qubits, i.e. coupling two or more transmon qubits or alternatively, coupling fluxonia. Exceptions where different types of qubits are coupled are, for example, Refs.~\cite{ku2020, noguchi2020, maiani2022}. In this paper, we consider the idea of using chips with heterogeneous qubits.  In particular we analyze how to realize microwave-activated two-qubit gates between capacitively coupled transmons and fluxonia. For fluxonia with medium frequencies, between $0.25$-$1.0 \, \mathrm{GHz}$, we show that the CR gate activated by driving the fluxonium at the transmon frequency is an ideal two-qubit gate candidate. In order to achieve similar gate times, the coupling capacitance needs to be larger as compared to the transmon-transmon case \cite{magesan2020}, but smaller than the fluxonium-fluxonium case \cite{dogan2022} due to the transmon being the better antenna.
A purely-capacitive coupling is easier to engineer than the inductive (combined with a small capacitive) coupling proposed in \cite{nguyen2022}. 

By means of a Schrieffer-Wolff analysis \cite{BRAVYI20112793}, we show that the CR effect is mainly mediated by the higher levels of the fluxonium and that it stays large over a wide range of transmon frequencies. Importantly, the frequency of the $\ket{1}- \ket{2}$ and $\ket{0}-\ket{3}$ transitions of the fluxonium should be designed to be relatively far away from the transmon frequency in order to limit residual, static $ZZ$ interactions and leakage in the fluxonium during the gate operation. By means of numerical simulations, which include noise, we show that the CR gate can be realized with leakage below $10^{-4}$ and gate fidelity above $99 \%$ with gate times between $100$ and $200 \, \mathrm{ns}$. On the other hand, we show that when the fluxonium frequency decreases to around $10 \, \mathrm{MHz}$ the CR effect vanishes. In this case, we find that a possible way of implementing a CPHASE gate is to drive to the higher levels of the fluxonium, similar to Ref.~\cite{ficheux2021}. The entangling power of the gate is then due to the coupling-induced hybridization between the bare $\ket{13}_0$ and $\ket{04}_0$ levels of the transmon-fluxonium system. We argue that despite the drawback of using the higher levels of the fluxonium, which have coherence times comparable to that of the transmon, the CPHASE gate can be implemented in $100$ to $200 \, \mathrm{ns}$ with arbitrary conditional phases and fidelities above $99 \%$.    

This paper is organized as follows: In ~\cref{sec:ft_sys} we introduce the transmon-fluxonium system. \cref{sec:crgate} presents the CR gate between a transmon and medium-frequency fluxonium. We provide a comparison with the transmon-transmon CR gate, highlighting the advantage of the transmon-fluxonium case with respect to the frequency crowding problem. We substantiate our understanding by explicit numerical simulations. In ~\cref{sec:cphase} we study a low-frequency fluxonium coupled to a transmon and propose a CPHASE gate similar to the one implemented in Ref.~\cite{ficheux2021} between two fluxonia. Also in this case, we perform numerical simulation of the gate showing that, by changing the pulse parameters, CPHASE gates with arbitrary conditional phases can be implemented. \cref{sec:arch} presents two possible transmon-fluxonium surface-code-like architectures based on either the CR or the CPHASE gate. We also perform a yield fabrication analysis for the CR gate architecture, similar to those in Ref.~ \cite{Hertzberg2021, morvan2021} for the transmon-transmon case and in Ref.~\cite{nguyen2022} for the fluxonium-fluxonium case. We conclude in \cref{sec:conclusions}.

 \section{The Transmon-Fluxonium system}
\label{sec:ft_sys}
\begin{figure}
\vspace{0.5 cm}
\centering
\includegraphics[width=0.48 \textwidth]{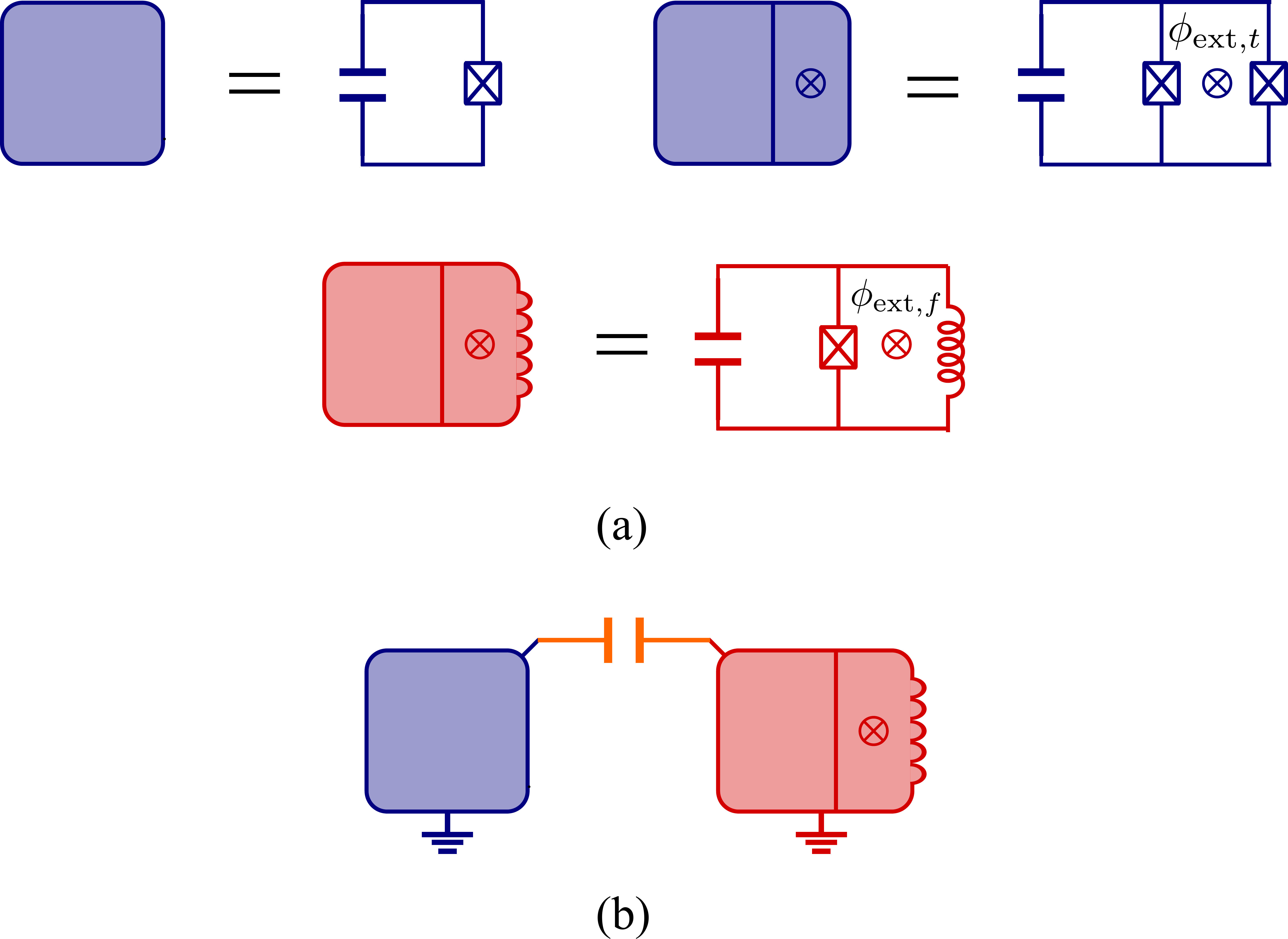}
\caption{(a) Symbolic representation of a fixed-frequency transmon (top left), a flux-tunable transmon (top right) and a fluxonium (bottom). (b) Capacitively coupled (fixed-frequency) transmon and fluxonium qubits.}
\label{fig:tr_coup_flx}
\end{figure}

\begin{figure}
\vspace{0.5 cm}
\centering
\includegraphics[width=0.45 \textwidth]{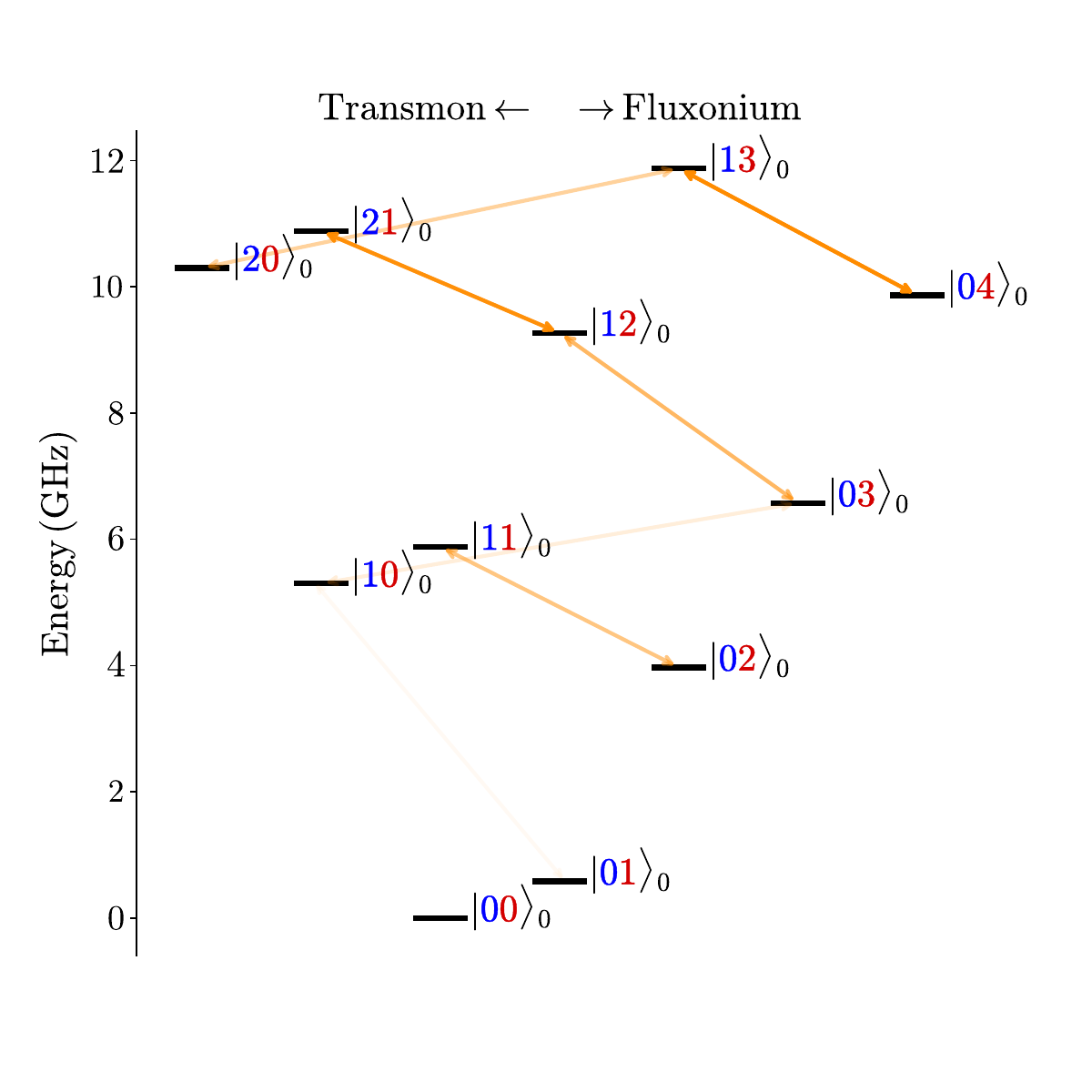}
\caption{An example of a typical energy level diagrams for a capacitively coupled transmon-fluxonium system. The arrows denote the levels that show non-zero matrix elements of the coupling Hamiltonian $\jc \qt \qf$: the darker the color of the arrow the larger the matrix element. The figure corresponds to the parameter set CR in Table \ref{tab:par_set} with $\omt/2 \pi = 5.3 \, \mathrm{GHz}$ and $\phi_{\mathrm{ext},f}=\pi$.}
\label{fig:tr_flx}
\end{figure}

The basic circuit of the coupled transmon-fluxonium system is shown in Fig.~\ref{fig:tr_coup_flx}b. 
Following standard circuit quantization \cite{vool2017, rasmussen2021} and directly approximating the transmon as a Duffing oscillator, the Hamiltonian can be written as
\begin{multline}
\label{eq:h}
H = \hbar \omt b^{\dagger}b + \hbar \frac{\anharm}{2} b^{\dagger} b^{\dagger}b b \\ + 4 \ecf  \qf^2 + \frac{1}{2}\elf \phif^2 - \ejf \cos\left(\phif - \phi_{\mathrm{ext},f}\right) \\ + \jc \qt \qf,
\end{multline}
with $\ecf, \elf, \ejf$ the fluxonium charging, inductive and Josephson energy respectively, $\omt/2 \pi$ the fundamental transmon frequency and $\anharm/2 \pi <0$ its anharmonicity. The transmon operators $b$ and $b^{\dagger}$ satisfy commutation relations $[b, b^{\dagger}] = I$, while the fluxonium (dimensionless) reduced charge $\qf$ and reduced flux $\phif$ operators satisfy $[\phif, \qf] = i I$, with $I$ the identity. The transmon charge operator $\qt$ can be expressed in terms of $b$ and $b^{\dagger}$ as 
\begin{equation}
\qt =  i \underbrace{\biggl( \frac{\ejt}{32 \hbar| \anharm |} \biggr)^{1/4}}_{\qzpf}(b^{\dagger} - b),
\end{equation}
where the transmon Josephson energy $\ejt$ is related to the qubit energy and anharmonicity by
\begin{equation}
\ejt = \frac{\hbar (\omt - \anharm )^2}{8 |\anharm|}.
\end{equation}
The coefficient $\qzpf$ represents the charge zero-point fluctuations of the transmon. In the usual Duffing approximation, the charging energy of the transmon is simply $\ect = \hbar |\anharm|$.

\begin{table*}[t]
  \centering
\renewcommand{\arraystretch}{1.8}
  \begin{tabular}{c | P{1.7cm}| P{1.7cm} | P{1.7cm} | P{1.7cm} |P{1.4cm}|  P{1.4cm}| P{1.6cm} | P{1.2cm} | P{1.4cm}}
  & \multicolumn{6}{c|}{Fluxonium} & \multicolumn{2}{c|}{Transmon} & \\
  \hline
    Parameter set & $\frac{\ecf}{h} \, (\mathrm{GHz})$ &  $\frac{\elf}{h} \, (\mathrm{GHz})$ & $\frac{\ejf}{h} \, (\mathrm{GHz})$  & $\frac{\omf{01}}{2 \pi} \, (\mathrm{MHz})$ &$T_{1}^{1 \mapsto 0} \, (\mathrm{\mu s})$  & $T_{1}^{3 \mapsto 0} \, (\mathrm{\mu s})$ & $\frac{\omt}{2\pi} \, (\mathrm{GHz})$ & $T_{1} \, (\mathrm{\mu s})$ & $\frac{\jc}{h} \, (\mathrm{MHz})$ \\
 
 \hline
    CR & $1.0$ & $1.0$ & $4.0$ & $582$ &$126$ & $20$ & $\in [4.2, 5.8]$ & $\approx 130 $ & $20$ \\
  \hline
  CPHASE & $1.0$ & $0.5$ & $8.0$ & 30 & $3700$ & $7$ & $4.37$ & $130$ & $30$ 
    
  \end{tabular}
  \caption{Parameter sets used in the manuscript. The transmon always has anharmonicity $\anharm/2\pi = -300 \, \mathrm{MHz}$. The relaxation times correspond to dielectric losses as described in Appendix \ref{app:error_model}. The dielectric loss tangent for the transmon is taken to be $\tan \delta_{\mathrm{diel},t}=3 \times 10^{-7}$ and assumed to be frequency independent. For the fluxonium, similar as in Ref.~\cite{nguyen2019}, we take a frequency-dependent dielectric loss tangent $\tan \delta_{\mathrm{diel},f}(\omega)=3.5 \times 10^{-6} (\omega/\omega_{\mathrm{ref}})^{0.15}$ with $\omega_{\mathrm{ref}}/2 \pi = 6.0 \, \mathrm{GHz}$. This is needed to take into account the various frequencies that are present in the fluxonium. The temperature of the environment is always assumed to be $T = 20 \, \mathrm{mK}$. Additional relaxation and excitation times for other relevant fluxonium transitions are reported in Table~\ref{tab:t1_flx} in ~\cref{app:error_model}.}
  \label{tab:par_set}
\end{table*}

In what follows we will assume the fluxonium to be biased with a reduced external flux $\phi_{\mathrm{ext}, f}=\pi$ so that it is operated in the double-well potential configuration. We will denote by $\ket{kl}_0$ the bare, uncoupled energy levels of the two qubits, where the first label $k$ identifies the transmon level, while the second label $l$ the fluxonium level. The symbol $\ket{kl}$ denotes the dressed, coupled energy level obtained by adiabatic continuation of the bare level $\ket{kl}_0$ when $\jc$ goes from $0$ to a nonzero value. The computational basis is defined as the dressed basis and the projector onto the computational subspace equals $P_c = \sum_{k, l=0}^1 \ket{kl}\bra{kl}$. The projector onto the leakage subspace is then $P_l=I - P_c$. We define $\omf{kl}$ as the transition frequency between the bare fluxonium levels $k$ and $l$ (not to be confused with $\omega_{kl}$ in, say, Eq.~\eqref{eq:cphase_cond} which denote the dressed energy levels of the transmon-fluxonium system). Also, let 
\begin{equation}
\label{eq:qfel}
\qfel{k > l} \,= \mathrm{Im}(\bra{k} \qf \ket{l}_f) = -i \bra{k} \qf \ket{l}_f, 
\end{equation}

be the imaginary part of the fluxonium matrix element with respect to the bare levels. 
Note that $\bra{k} \qf \ket{k}_f=0$. In addition, since at $\phi_{\mathrm{ext}, f}=\pi$ the fluxonium Hamiltonian, just like the transmon, has a parity symmetry, we also have $\bra{k} \qf \ket{k+2 m}_f=0$ for $m \in \mathbb{N}$.  

\begin{figure}
\vspace{0.5 cm}
\centering
\includegraphics[width=0.49 \textwidth]{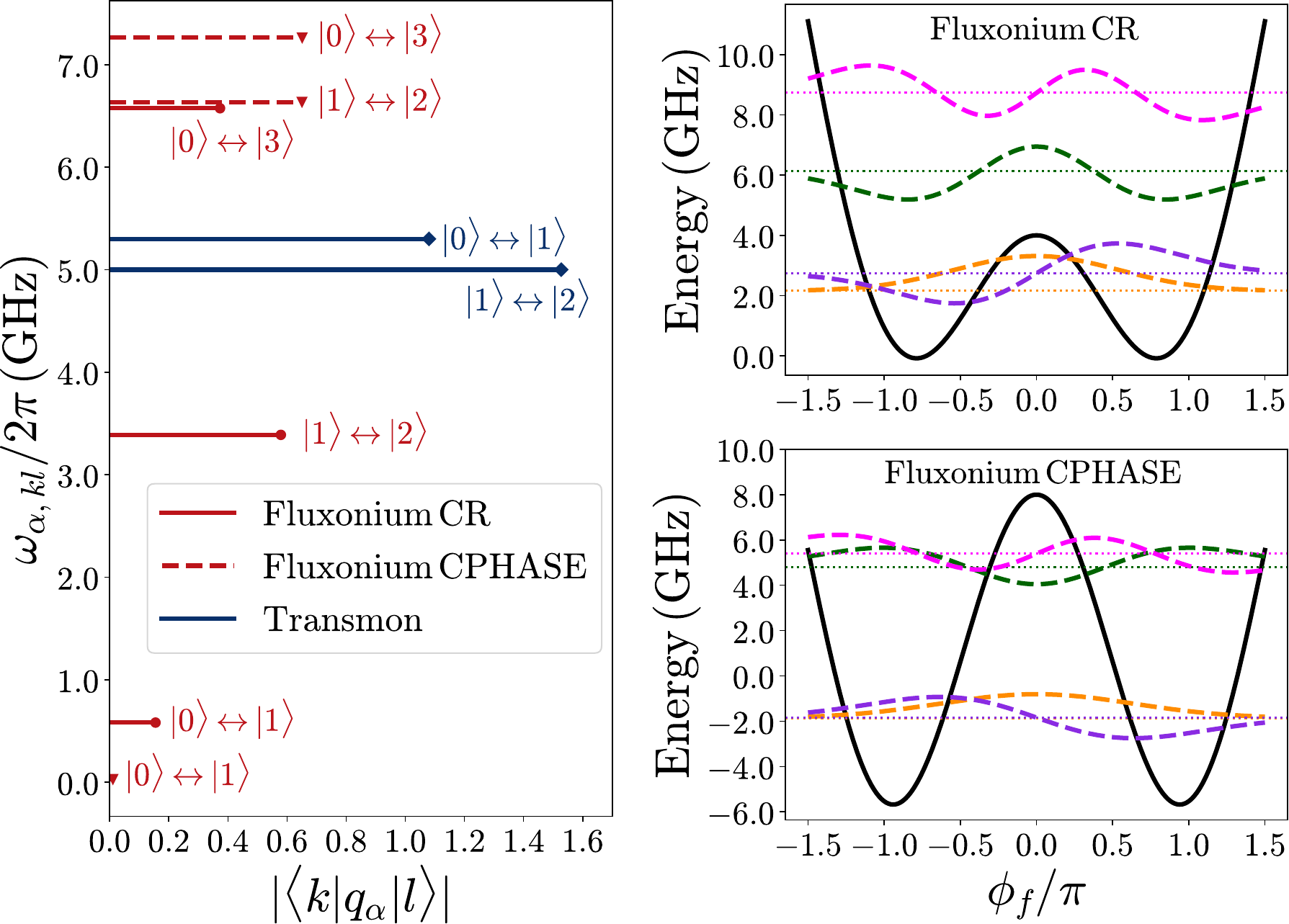}
\caption{(Left) Matrix elements of the fluxonium charge operator ($\alpha=f$) for the fluxonia in Table \ref{tab:par_set} and of the transmon charge operator ($\alpha=t$) with corresponding transition frequency on the $y$ axis. The transmon is taken to have fundamental frequency $\omt/2 \pi = 5.3 \, \mathrm{GHz}$ and anharmonicity $\anharm/2 \pi = -300 \, \mathrm{MHz}$. (Right) First $4$ eigenfunctions and potential energy for the fluxonia with parameter set CR (top) and CPHASE (bottom) in Table \ref{tab:par_set}. The dotted lines represent the energy corresponding to each level.}
\label{fig:flx_mat_elem}
\end{figure}

For the parameters listed in Table \ref{tab:par_set}, the energy levels $\ket{02}_0$ and $\ket{03}_0$ have frequencies of the same order of magnitude as the $\ket{10}_0, \ket{11}_0$ levels as seen in the level diagram in Fig.~\ref{fig:tr_flx}. Due to the relatively large matrix elements of the two lowest levels of the fluxonium with the higher levels (see Fig.~\ref{fig:flx_mat_elem}), the $\qf \qt$ term in the Hamiltonian directly couples levels $\ket{10}_0 \leftrightarrow \ket{03}_0$ and levels $\ket{11}_0 \leftrightarrow \ket{02}_0$. As we show in Appendix \ref{app:sw}, the coupling to these levels induces a spurious $ZZ$ coupling, but also gives rise to the CR interaction.

The gates considered in this paper will be activated by a microwave drive on the fluxonium. A drive at carrier frequency $\omd$ can be modelled by the following time-dependent Hamiltonian
\begin{equation}
\label{eq:hdrive}
H_{\mathrm{drive}}(t) = \hbar g(t)\epsd \cos(\omd t + \phased) \qf,
\end{equation} 
with $0 \le g(t) \le 1$ a dimensionless envelope function and $\epsd$ the maximum drive amplitude that characterizes the drive strength and $\phased$ the phase of the drive.

\label{sec:crgate}
\begin{figure}
\centering
\includegraphics[width=0.4\textwidth]{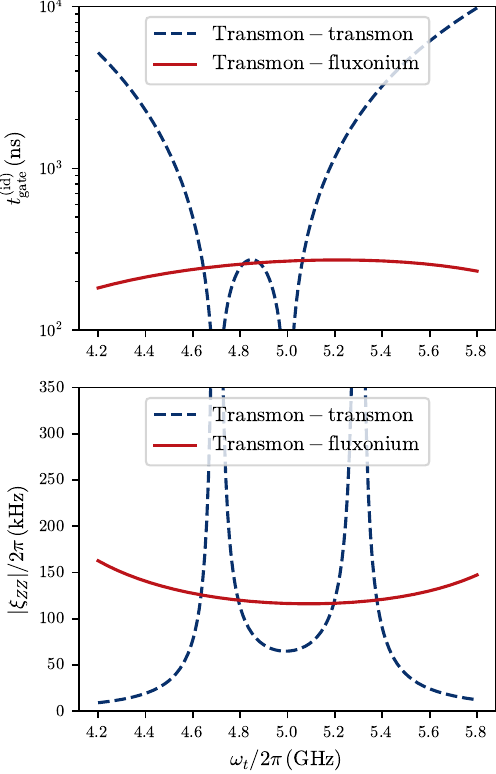}
\caption{Comparison of the gate time (top) and the residual, static $ZZ$ coupling (bottom) for the transmon-transmon and the transmon-fluxonium CR gate. The frequency of the control transmon in the transmon-transmon case is fixed to $\omega_c/2 \pi =5.0 \, \mathrm{GHz}$ and both transmons have anharmonicity $\anharm/2 \pi =\delta_c/2\pi= -300 \, \mathrm{MHz}$. 
The coupling between the transmons is set to $\jc/h = 2 \, \mathrm{MHz}$. In the transmon-fluxonium case the parameters are taken as in parameter set CR in Table \ref{tab:par_set}. In the top figure the drive on the control transmon in the transmon-transmon case is set to $\epsd/2 \pi = 30 \, \mathrm{MHz}$, while the drive on the control fluxonium in the transmon-fluxonium case is taken to be $\epsd/2 \pi = 300 \, \mathrm{MHz}$. We use Eqs.~\eqref{eq:mu_cr_sw_mt} and \eqref{eq:mu_cr_tt_sw} to estimate the ideal gate time in Eq. \eqref{eq:gate_time_cr}. The $ZZ$ coupling $\xi_{ZZ}$ is evaluated via numerical diagonalization using Eq.~\eqref{eq:zz_coup_def}. For the transmon-transmon case we see the $ZZ$ coupling blowing up at the resonances $\ket{11} \leftrightarrow \ket{02}$ and $\ket{11}\leftrightarrow \ket{20}$ (not at the resonance $\ket{01}\leftrightarrow \ket{10}$!), while the gate time blows up at the resonances $\ket{01}\leftrightarrow \ket{10}$ and $\ket{11} \leftrightarrow \ket{02}$, leaving a narrow frequency window of opportunity.}
\label{fig:cr_comparison}
\end{figure}
\section{The Cross-Resonance gate for medium frequency fluxonia}
The CR effect manifests itself when we drive one of the qubits (the control) at the fundamental frequency of the other (the target). The effect arises due to the presence of the coupling term, which enables the drive operator of the control qubit to drive transitions between energy levels of the target qubit. Importantly, the drive on the target qubit depends on the state of the control. Here we take the fluxonium as the control qubit and the transmon as the target qubit, so we drive the fluxonium at the fundamental frequency of the transmon. This choice is motivated by the fact that we can have large CR effect leading to a gate time around $100 \, \mathrm{ns}$, while the same does not hold if we take the transmon as control and the fluxonium as target. At low drive strengths \cite{magesan2020, krantz2019}, the CR effect can be simply understood by looking at the matrix elements of the charge operator in the dressed basis. We provide a perturbative analysis of the CR coefficient in Appendix \ref{app:sw}.

Including the envelope function, the fundamental CR Hamiltonian is
\begin{equation}
H_{\mathrm{CR}}(t) = \hbar g(t) \mu_{\mathrm{CR}} X_{t} Z_f,
\end{equation}
with $X_t$ the transmon Pauli $X$ operator and $Z_f$ the fluxonium Pauli $Z$ operator. As remarked in Appendix \ref{app:sw}, by changing the phase of the drive we can always make $\mu_{\mathrm{CR}}$ positive. The CR gate with the fluxonium as control and the transmon as target is given by the unitary
\begin{equation}
U_{\mathrm{CR}} = e^{-i \frac{\pi}{4} X_t Z_f},
\end{equation}
and thus, it is implemented by letting $H_{\mathrm{CR}}(t)$ act for a time $t_{\mathrm{gate}}$ such that
\begin{equation}
\label{eq:cr_condition}
\frac{\mu_{\mathrm{CR}}}{\hbar} \int_0^{t_{\mathrm{gate}}} d t g(t) = \frac{\pi}{4}.
\end{equation}
If $g(t)$ is a constant equal to $1$ we obtain the simple formula
\begin{equation}
\label{eq:gate_time_cr}
t_{\mathrm{gate}}^{(\mathrm{id})} = \frac{\hbar \pi}{4 \mu_{\mathrm{CR}}}.
\end{equation}
Using additional single-qubit gates the CR gate can be turned into a CNOT gate \cite{krantz2019}.

In Appendix \ref{app:sw} we derive an approximate formula for the CR coefficient in the transmon-fluxonium case, using a second-order Schrieffer-Wolff transformation, which reads
\begin{equation}
\label{eq:mu_cr_sw_mt}
\mu_{\mathrm{CR}} = \frac{\jc \qzpf}{4 \hbar} \biggl[\frac{\qfel{3}{0}^2}{\omt - \omf{30}} - \frac{\qfel{2}{1}^2}{\omt -\omf{21} } \biggr] \epsd.
\end{equation}
This expression has to be compared with the standard expression for the transmon-transmon case with control qubit $c$ and target qubit $t$ \cite{magesan2020, juelich_le}
:
\begin{equation}
\label{eq:mu_cr_tt_sw}
\mu_{\mathrm{CR}}^{(tt)} = -\frac{\jc q_{\mathrm{zpf}, t} q_{\mathrm{zpf}, c}^2}{\hbar(\omega_c - \omega_t)} \biggl[\frac{\delta_c}{\delta_c + \omega_c -\omt} \biggr] \frac{\epsd}{2},
\end{equation}
where $q_{\mathrm{zpf}, t}$ ($q_{\mathrm{zpf}, c}$), $\omt/2 \pi$ ($\omega_c/2 \pi$)
are the charge zero point fluctuation and the fundamental frequency of the target (control) transmon, respectively, and $\delta_c/ 2 \pi$ the anharmonicity of the control transmon. 

 We show a comparison between the transmon-transmon and transmon-fluxonium CR gate in Fig. \ref{fig:cr_comparison}. Due to the small anharmonicity of the transmons, Eq.~\eqref{eq:mu_cr_tt_sw} predicts that the CR coefficient is only large enough to lead to a gate time below $300 \, \mathrm{ns}$ when the condition $\omega_c + \delta_c \le \omega_t \le \omega_c$ is satisfied, see Fig.~\ref{fig:cr_comparison} (top). Outside this region the gate time quickly increases to values above $1 \, \mu s$. Thus, the CR gate for transmon-only systems requires careful frequency engineering and the constraints leads to frequency crowding \cite{Hertzberg2021,morvan2021}. 
 
In contrast, the transmon-fluxonium CR gate can be activated for a large range of target transmon frequencies and presents more stable CR gate times. While a larger drive strength $\epsd$ is necessary for the transmon-fluxonium case in order to compensate for the smaller charge matrix elements of the fluxonium, we show below, via numerical simulations, that small leakage and high-fidelities can be achieved. 
In Fig.~\ref{fig:cr_comparison} (bottom) we also plot the residual, static $ZZ$ coupling defined as
\begin{equation}
\label{eq:zz_coup_def}
\frac{\xi_{ZZ}}{2 \pi} = \frac{E_{11} - E_{10} - E_{01} + E_{00}}{h},
\end{equation}
with $E_{kl}$ the dressed eigenenergy of level $\ket{kl}$. While the transmon-transmon case achieves a smaller minimum $ZZ$ coupling, the transmon-fluxonium case shows a more stable $\xi_{ZZ}$ without resonant peaks.

\begin{figure}
\centering
\includegraphics[width=0.4 \textwidth]{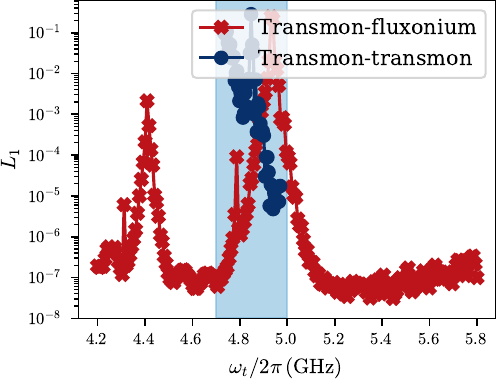}
\caption{Average leakage for a simulated, noiseless CR gate for the transmon-transmon and the transmon-fluxonium case as a function of the target frequency. In both cases parameters are taken as in Fig.~\ref{fig:cr_comparison} and we set $\omd = \omt$. We numerically simulate the gates using a piece-wise Gaussian envelope (see Appendix \ref{app:pulse}) with rise time $t_{\mathrm{rise}} = 10 \, \mathrm{ns}$ and gate time chosen to satisfy the CR condition in Eq.~\eqref{eq:cr_condition}. We plot the result for the transmon-transmon case only in the relevant, blue-shaded region such that $\omega_c + \delta_c \le \omt \le \omega_c$. In the simulations we include $3$ bare levels for the transmons and $8$ for the fluxonium.}
\label{fig:leak_comparison}
\end{figure}

In addition, we observe that the average leakage $L_1$ (see~\cref{app:fid} for a definition) in the CR gate for the transmon-fluxonium case is generally lower than for the transmon-transmon case for different target transmon frequencies, see Fig.~\ref{fig:leak_comparison}. In the transmon-fluxonium case, and in the frequency range $4.2-5.8 \, \mathrm{GHz}$, we identify two peaks in the leakage at $\omt/2 \pi \approx 4.41 \, \mathrm{GHz}$ and at $\omt/2 \pi \approx 4.93 \, \mathrm{GHz}$. The former is due to a three-photon transition between the levels $\ket{0}$ and $\ket{5}$ of the fluxonium with frequency $\omf{05}/2 \pi = 13.23 \, \mathrm{GHz}$, while the latter is caused by a two-photon transition between levels $\ket{0}$ and $\ket{4}$ of the fluxonium with frequency $\omf{04}/2 \pi = 9.86 \, \mathrm{GHz}$. These are frequency collisions that one must avoid in order for the CR gate to achieve high fidelities (see also the discussion in Sec.~\ref{sec:arch}). Away from these frequencies the leakage can be as low as $10^{-7}$ and this happens for a wide target frequency range. In contrast, the transmon-transmon case achieves leakage below $10^{-5}$ only close to $\omt/2\pi = \omega_{c}/2 \pi = 5.0 \, \mathrm{GHz}$, where however a problem of qubit addressability emerges. Moreover, the leakage increases when the two-photon transition between levels $\ket{0}$ and $\ket{2}$ of the control is triggered at $\omt/2 \pi = 4.85 \, \mathrm{GHz}$ and when $\omt$ matches the $\ket{1}$-$\ket{2}$ transition of the control at $\omt/2\pi = 4.7 \, \mathrm{GHz}$.

\begin{figure}
\centering
\includegraphics[width=0.49 \textwidth]{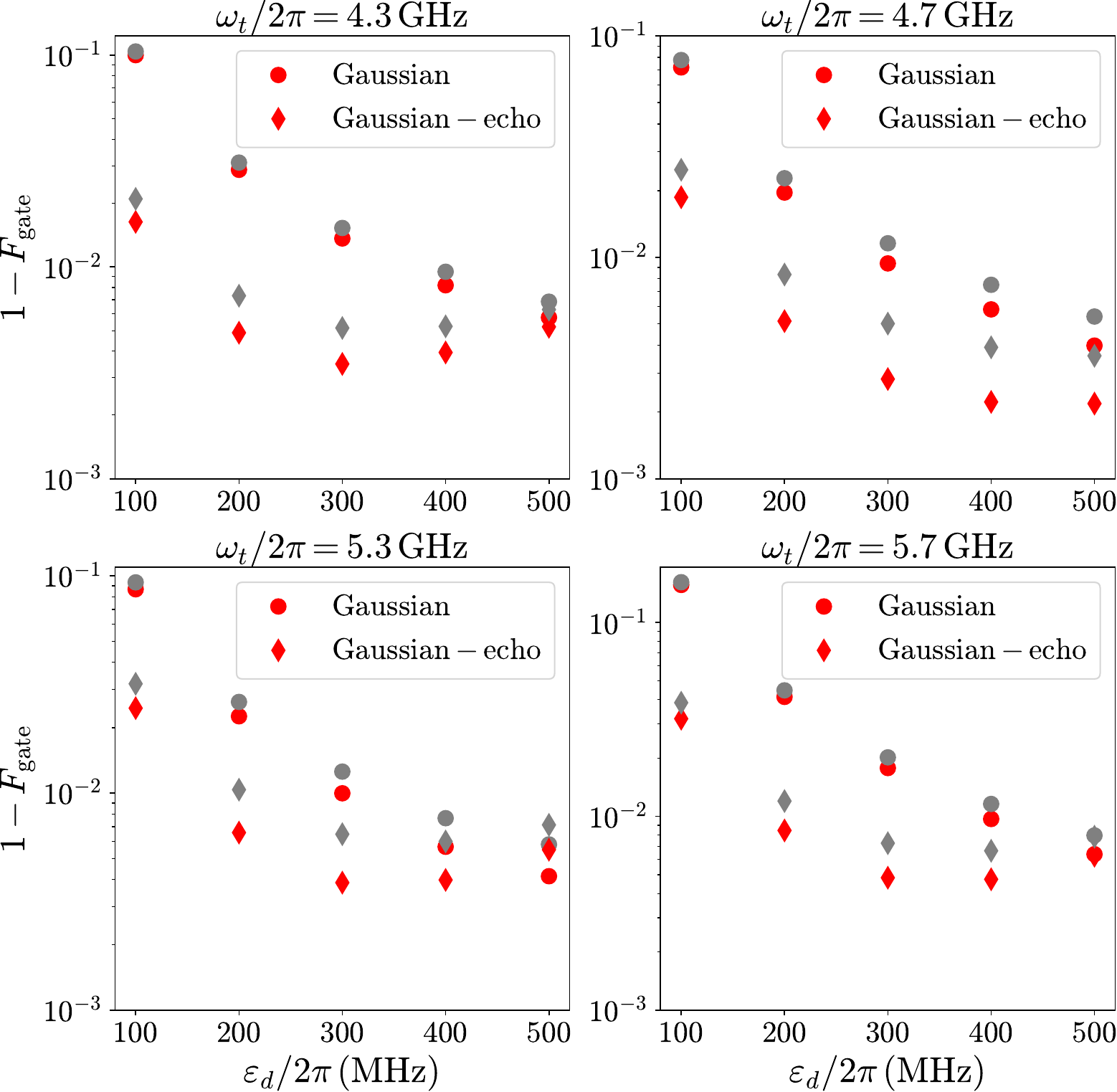}
\caption{Gate infidelity as a function of the drive strength for four different target transmon frequencies in the transmon-fluxonium case. The piece-wise Gaussian envelope are always taken to have rise time $t_{\mathrm{rise}} = 10 \, \mathrm{ns}$. We also plot in gray the infidelities in the presence of dielectric loss. When $\epsd/ 2 \pi = 300 \, \mathrm{MHz}$, gate times for the simple Gaussian pulse (no echo) are $t_{\mathrm{gate}} = 233, 297, 330, 295 \, \mathrm{ns}$ for $\omt/2 \pi = 4.3, 4.7, 5.3, 5.7 \, \mathrm{GHz}$, respectively. In the simulations we include $3$ bare levels for the transmons and $6$ for the fluxonium.}
\label{fig:fid_drive}
\end{figure}

In Fig.~\ref{fig:fid_drive} we plot gate infidelities in the noiseless and noisy case as a function of the drive strength for the four target transmon frequencies that we consider in the yield analysis in  Sec.~\ref{sec:arch}. We refer the reader to~\cref{app:fid} for the definition of gate fidelities and to~\cref{app:error_model} for details of the dielectric loss error model used for the noisy simulations. We consider a simple piece-wise Gaussian pulse and its echo version, where two piece-wise Gaussian pulses on the fluxonium at frequency $\omd/2\pi=\omt/2\pi$ and with opposite phase, are interleaved with single-qubit $\pi$ rotations around the $X$ axis on the fluxonium qubit (as for the transmon-transmon case \cite{corcoles:RB}). We provide more details about the pulses in~\cref{app:pulse}. We see that for the chosen target frequency range, the echo pulse generally outperforms the simple Gaussian pulse, especially at low drive strengths. This is because the echo pulse ideally cancels the effect of the unwanted $ZZ$ interaction (besides canceling the $X$-rotation on the transmon qubit), while the smaller the drive strength the larger the effect of the $ZZ$ interaction is compared to the CR effect~\cite{malekakhlagh2020}, which explains the lower fidelities in the simple pulse case. In all cases, we observe that with the echo pulse gate fidelities can be at least $99.5 \% $ in the noiseless case and at least $99.3 \%$ in the noisy case for all target frequencies. We remark that these were obtained without any pulse optimization and simply by matching the CR condition in Eq.~\eqref{eq:cr_condition} and the corresponding one for the echo pulse (see~\cref{app:pulse}). Optimal control techniques developed for the CR gate~\cite{kirchhoff2018} and
more detailed techniques to understand the sources of error~\cite{korotkov2019, malekakhlagh2022} for the transmon-transmon case can also be applied to the transmon-fluxonium case to achieve higher fidelities.

\section{The CPHASE gate for low frequency fluxonia}
\label{sec:cphase}
If we keep the fluxonium charging energy $E_{C,f}$ fixed and decrease the ratio between $\elf/\ejf$, the energy barrier between the two lowest minima of the fluxonium potential increases, while the kinetic term stays constant. In this limit, the two lowest eigenstates of the fluxonium are an even and odd superpositions of two states that are more and more localized in the left and right well of the potential, respectively, see Fig.~\ref{fig:flx_mat_elem}(b). As a result, the fundamental frequency of the fluxonium, i.e., the energy splitting between the two lowest levels, decreases, as well as the magnitude of the matrix elements $\bra{0}\qf \ket{1}$, $\bra{0}\phif \ket{1}$ of the charge and flux operators between the two lowest levels, see Fig.~\ref{fig:flx_mat_elem}(a). This naturally leads to longer coherence times, although the control of the quantum state of the fluxonium becomes more involved \cite{zhang2021}. In this scenario, the frequencies of the transmon and of the fluxonium are extremely different, since the fundamental frequency of the fluxonium can be smaller than $100 \, \mathrm{MHz}$.

\begin{figure}
\centering
\includegraphics[width=0.45 \textwidth]{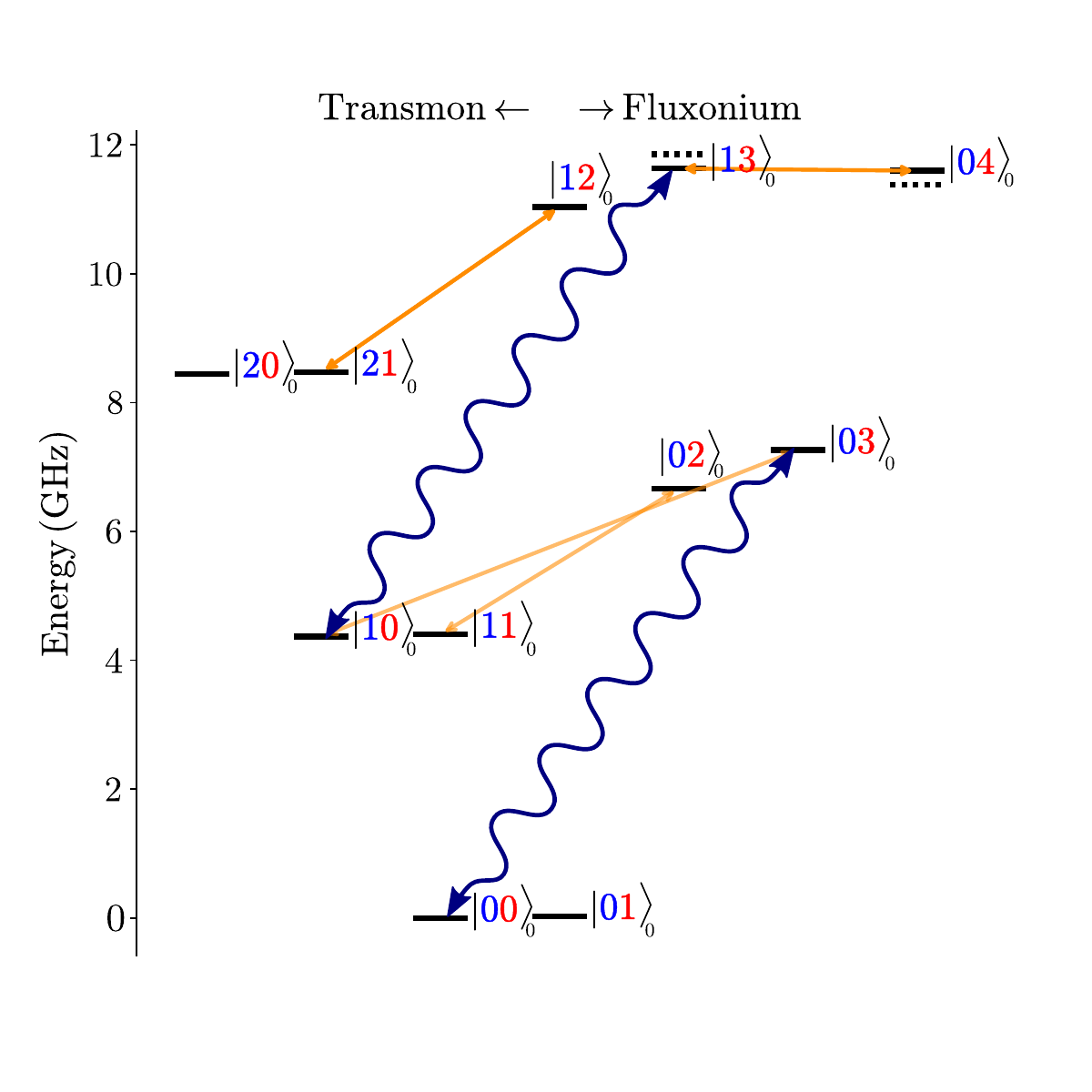}
\caption{Energy level diagram for a low frequency fluxonium coupled to a transmon corresponding to parameter set CPHASE in Table \ref{tab:par_set}. The blue wavy lines identify the levels that are driven. The dashed lines above the level $\ket{13}_0$ and below the level $\ket{04}_0$ represent the dressed levels $\ket{13}$ and $\ket{04}$, respectively.}
\label{fig:en_lev_low}
\end{figure}

Moreover, the CR scheme that we analyzed in Sec.~\ref{sec:crgate} is not applicable in this case because of a vanishing cross-resonance coefficient $\mu_{\mathrm{CR}}$. We can understand this by looking at the approximate, analytical formula for $\mu_{\mathrm{CR}}$ in Eq.~\eqref{eq:mu_cr_sw_mt}. In the limit of low fluxonium frequency the difference between the frequencies $\omf{30}$ and $\omf{21}$ becomes smaller (see the energy levels in Fig.~\ref{fig:en_lev_low} for an example). In order to have a large $\mu_{\mathrm{CR}}$ the two terms on the right-hand side of Eq.~\eqref{eq:mu_cr_sw_mt} need to constructively interfere, i.e., the transmon frequency must be chosen between $\omf{30}$ and $\omf{21}$, leading to a small range of transmon frequencies with a sizable cross-resonance coefficient. Outside this transmon frequency range the $\mu_{\mathrm{CR}}$ goes to zero also because $|\qfel{2}{1}| \approx |\qfel{3}{0}|$ in the low fluxonium frequency limit.

In this section we show that the scheme proposed in Ref.~\cite{ficheux2021} to realize a CPHASE gate between two fluxonia can be adapted using different fluxonium levels. In Fig.~\ref{fig:en_lev_low} we show the energy level diagram for parameter set CPHASE which we will use in this section, see Table~\ref{tab:par_set}. We see that the system is chosen to have (approximately) a resonance between the transmon frequency and the $\ket{3}-\ket{4}$ transition of the fluxonium, i.e., $\omt \approx \omf{34}$.
Denoting by $E_{kl}$ the dressed energies of level $\ket{kl}$ and by $E_{kl}^{(0)}$ the bare energy of level $\ket{kl}_0$, we will thus have $E_{13} - E_{04} \neq E_{13}^{(0)} - E_{04}^{(0)}$ due to the coupling term. This energy shift in turn gives rise to a difference between the frequencies associated with the $\ket{00}-\ket{03}$ transition and with the $\ket{10}-\ket{13}$ transition, expressed by a parameter $\Delta$:
\begin{equation}
\frac{\Delta}{2 \pi} = \frac{(E_{13}- E_{10}) - (E_{03} - E_{00})}{h} \neq 0.
\label{eq:defdelta}
\end{equation} 
For parameter set CPHASE in Table~\ref{tab:par_set} we have $\Delta /2 \pi = 14.0 \, \mathrm{MHz}$.
This effect can be used to implement a microwave-activated CZ gate and more generally a CPHASE gate with an arbitrary phase $\phi$ given by the general unitary
\begin{equation}
\label{eq:cphase}
\mathrm{CPHASE}(\phi) = \begin{pmatrix}
1 & 0 & 0 & 0 \\
0 & 1 & 0 & 0 \\
0 & 0 & 1 & 0 \\
0 & 0 & 0 & e^{i \phi}
\end{pmatrix}.
\end{equation}
As schematically shown in Fig.~ \ref{fig:en_lev_low}, the idea is to apply a drive on the fluxonium qubit with a frequency approximately given by the $\ket{0}-\ket{3}$ transition frequency. More precisely, the drive amplitude and the drive frequency is chosen so that the generalized Rabi frequencies of the $\ket{00}-\ket{03}$ and $\ket{10}-\ket{13}$ transitions are matched to a certain value $\Omega$, which is expressed in the condition
\begin{multline}
\label{eq:cphase_cond}
\sqrt{\epsd^2 \qfel{10}{-13}^2 + (\omega_{13}-\omega_{10} - \omd)^2} = \\ \sqrt{\epsd^2 \qfel{00}{-03}^2 + (\omega_{13}-\omega_{10} - \omd - \Delta)^2} = \Omega,
\end{multline}
where $\qfel{10}{-13}=|\bra{10}\qf\ket{13}|$ and $\qfel{00}{-03}= |\bra{00}\qf\ket{03}|$. 

This condition leads to drive frequencies that are between the frequencies of the $\ket{00}-\ket{03}$ and $\ket{10}-\ket{13}$ transition, which for parameter set CPHASE in Table \cref{tab:par_set} is 
$\omd/2 \pi \approx 7.26 \, \mathrm{GHz}$. Also, the typical drive strength is $\epsd/2 \pi \approx 10 \, \mathrm{MHz}$. 

Eq.~\eqref{eq:cphase_cond} guarantees that, assuming $g(t)=1$, after a time $t_{\mathrm{gate}}=2\pi/\Omega$ both transitions give rise to a Rabi oscillation which ideally induces no leakage. The nonzero $\Delta$ gives rise to a differential phase $\phi \approx \pi \Delta/\Omega$ which is acquired by the state $\ket{11}$, see the detailed analysis of the gate in Appendix \ref{app:cphase}. By means of single-qubit $Z$ rotations the implemented gate can be turned into the CPHASE gate in Eq.~\eqref{eq:cphase}. 

\begin{figure}
\centering
\includegraphics[width=0.4 \textwidth]{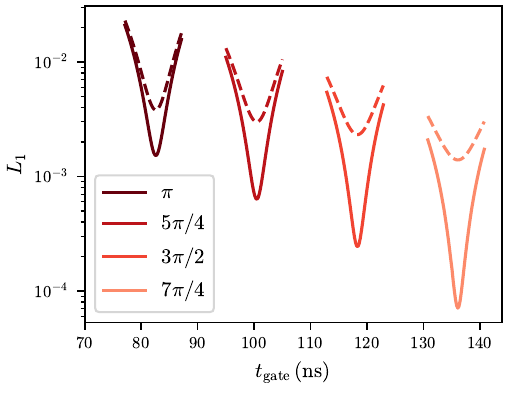}
\caption{Leakage as a function of gate time for different target conditional phases. We use a piece-wise Gaussian envelope $g(t)$ with rise time $t_{\mathrm{rise}}=10 \, \mathrm{ns}$. Drive frequencies and drive strengths are chosen to satisfy Eq.~\eqref{eq:cphase_cond} with $\Omega=\Delta/(\pi \phi)$ which depends on the target conditional phase $\phi$. Minimum leakage (approximately) corresponds to the condition $\int_{0}^{t_{\mathrm{gate}}} dt \,g(t) = 2\phi/\Delta$. Solid lines represent the coherent results, while the dashed lines represent the noisy ones in the presence of dielectric losses. In the simulations we include $3$ bare levels for the transmons and $5$ for the fluxonium.}
\label{fig:leakcphase}
\end{figure}

Crucial for the implementation of the gate is that the system undergoes a Rabi oscillations for both the involved transitions, inducing little leakage. We investigate this property in Fig.~\ref{fig:leakcphase}, where the average leakage $L_1$ (see its definition in Appendix \ref{app:fid}) is plotted as a function of gate time for different target conditional phases. We see that in the coherent case there is a very sharp minimum where the leakage is minimized. Also, the minimum value of the leakage decreases for larger conditional phases, which have larger optimal gate time since $t_{\rm gate} \approx 2\phi/\Delta$. 

However, comparing Figs.~\ref{fig:leak_comparison} and \ref{fig:leakcphase}, we notice that the leakage is much higher than that achievable with the CR gate discussed in Sec.~\ref{sec:crgate} due to the fact that the CPHASE gate is directly populating non-computational states during the gate. While lower leakage could be potentially achieved with optimized pulses, leakage is definitely more pronounced for this CPHASE gate as compared to the CR gate, assuming the CR gate is operated away from frequency collisions. In addition, in Fig.~\ref{fig:leakcphase} we observe (unsurprisingly) that the presence of noise increases the leakage compared to the noiseless case, and makes the leakage-minima less sharp.
Also, the use of higher levels of the fluxonium inevitably exposes the system to the shorter decay times of these levels. Despite these facts, we obtain optimal gate fidelities in the noise free case of $\{99.76 \%, 99.87 \%, 99.93 \%, 99.96 \% \}$ and in the noisy case of $\{99.36 \%, 99.44 \%, 99.51 \%, 99.61 \% \}$ for the conditional phases $\phi=\{\pi, 5 \pi/4, 2 \pi/2, 7\pi/4 \}$, respectively. The noise free gate fidelity is limited by leakage. In particular, in the noiseless case and when the condition Eq.~\eqref{eq:cphase_cond} is matched, the state with the highest probability of leakage is $\ket{04}$. This state accounts for approximately $95 \%$ of the leakage. Due to the hybridization of the bare levels $\ket{13}_0$ and $\ket{04}_0$, the dressed state $\ket{04}$ also acquires a non-zero fluxonium charge matrix element, and thus the drive can stimulate the $\ket{10}-\ket{04}$ transition, although it is more off-resonant than the $\ket{10}-\ket{13}$ transition.  This also explains the decrease of leakage with the increase of the target conditional phase in Fig.~\ref{fig:leakcphase}. In fact, the larger the target conditional phase, the larger the ideal gate time and the smaller the drive amplitude, which causes the unwanted $\ket{04}$ to be less populated. The remaining $5 \%$ of the leakage is due to the imperfect cancellation of the $\ket{13}$ population. We believe that these two main sources of leakage can be both reduced by further pulse optimization. Finally, we remark that in the noisy case we see an increase in the population of $\ket{13}$ as well as of $\ket{03}$. The latter seems due to a direct relaxation from the $\ket{04}$ state (see Table \ref{tab:t1_flx} in \cref{app:error_model} for the relaxation time).

When the drive is turned off, the static, residual $ZZ$ coupling is small and it is evaluated to be $40 \, \mathrm{kHz}$ for parameter set CPHASE in Table \ref{tab:par_set} used in this section. As can be seen from Fig.~\ref{fig:en_lev_low}, the bare levels $\ket{11}_0$ and $\ket{02}_0$ have a non-zero matrix element induced by the capacitive coupling. However, these levels are far detuned in frequency, by more than $2 \, \mathrm{GHz}$, and as a result have small level hybridization. The same holds for the pair of bare levels $\ket{10}_0, \ket{03}_0$.  As shown in Appendix \ref{app:sw}, these transitions are the main cause of $ZZ$ coupling in this system, which is suppressed given the large detuning between the levels.

\section{Architectures based on fluxonium and transmon}
\label{sec:arch}
In this section we provide some architectural considerations for multi-qubit transmon-fluxonium chips based on either the CR gate or the CPHASE gate analyzed in the previous sections. For concreteness, we focus on a surface code architecture, where each qubit is directly coupled to at most four neighbors \cite{versluis2017, Andersen2020, Marques2022}. In this case, either the transmons or the fluxonia play the role of the data or ancilla qubit in the surface code. For both the CR and CPHASE gate, we drive the fluxonium qubit and since up to four transmons couple to the same fluxonium qubit we need to examine how to avoid operation crosstalk. We then explore the expected yield for the architecture based on the CR gate, for which we expect frequency collisions to impact the fidelity of operations.

\subsection{Frequency allocation and operations}
\begin{figure}
\centering
\includegraphics[width=0.4 \textwidth]{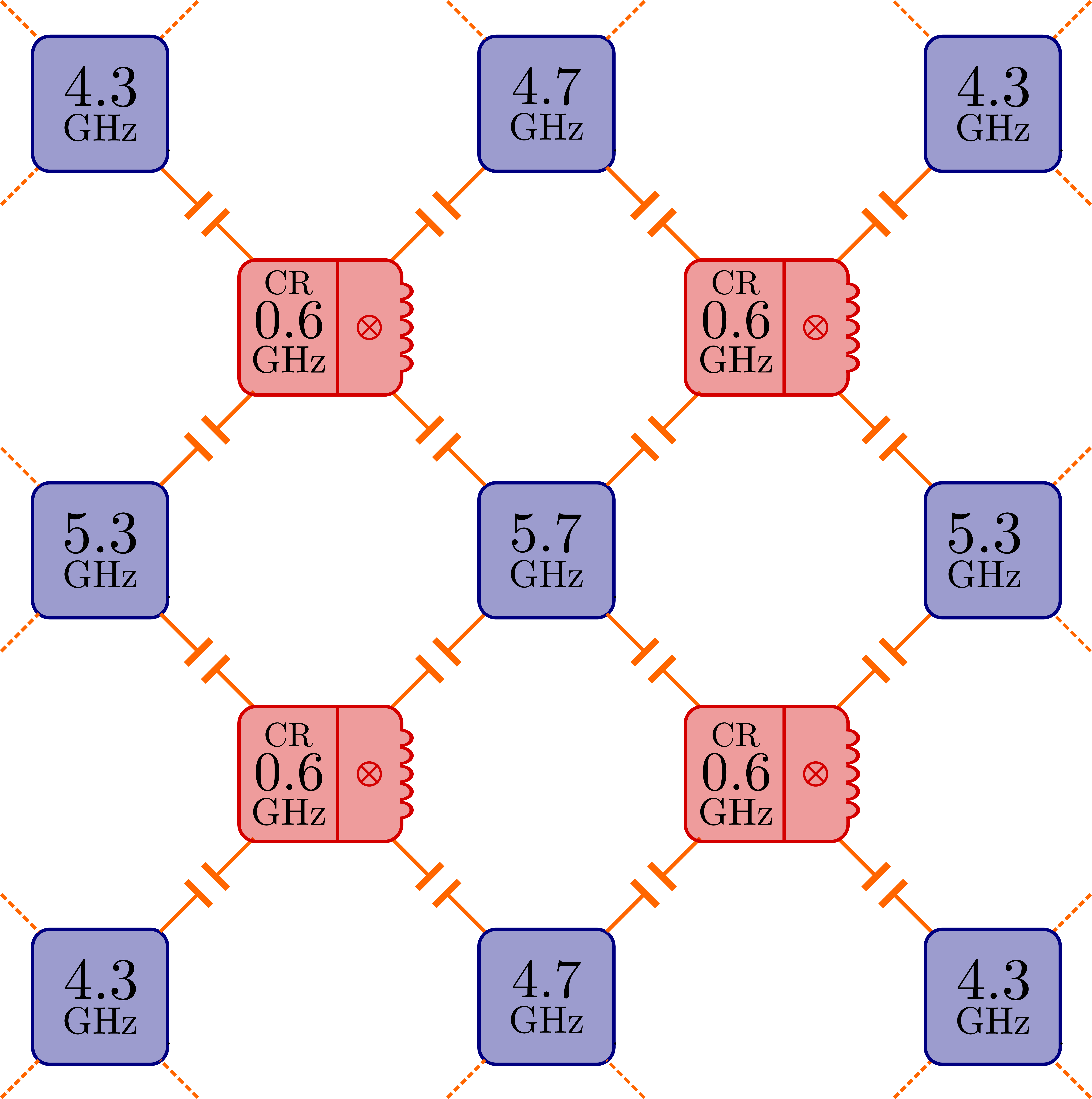}
\caption{Surface code archtitecture for fixed frequency transmons (blue) and  medium frequency fluxonia (red) based on the CR gate. We choose transmons at four different fundamental frequencies $\omt/2\pi = \{4.3, 4.7, 5.3, 5.7 \} \, \mathrm{GHz}$ and anharmonicity $\anharm/2 \pi = -0.3 \, \mathrm{GHz}$, while all the fluxonia have target parameters as in parameter set CR in Table \ref{tab:par_set} with fundamental frequency $\omf{01}/2 \pi \approx 0.6 \, \mathrm{GHz}$.}
\label{fig:arch_cr}
\end{figure}

For the CR gate we can make use of the wide range of transmon frequencies over which the gate can be implemented. A possible frequency setup is shown in Fig. \ref{fig:arch_cr}. In this case the architecture is fully microwave, without the need of frequency tunable transmons or tunable couplers, thus, flux control is only needed for static biasing of the fluxonia.

The fixed-frequency transmons are well-separated in frequency by at least $400\, \mathrm{MHz}$. With these parameters the gate times, assuming $\jc = 20 \, \mathrm{MHz}$ and $\epsd/2 \pi = 300 \, \mathrm{MHz}$, would be roughly between $200$ and $350 \, \mathrm{ns}$ and the residual $ZZ$ coupling between $100$ and $150 \, \mathrm{kHz}$. In this range of frequencies the $ZZ$ coupling is quite stable and does not have any sharp peaks (see Fig.~\ref{fig:cr_comparison}). 

\begin{figure}
\centering
\includegraphics[width=0.4 \textwidth]{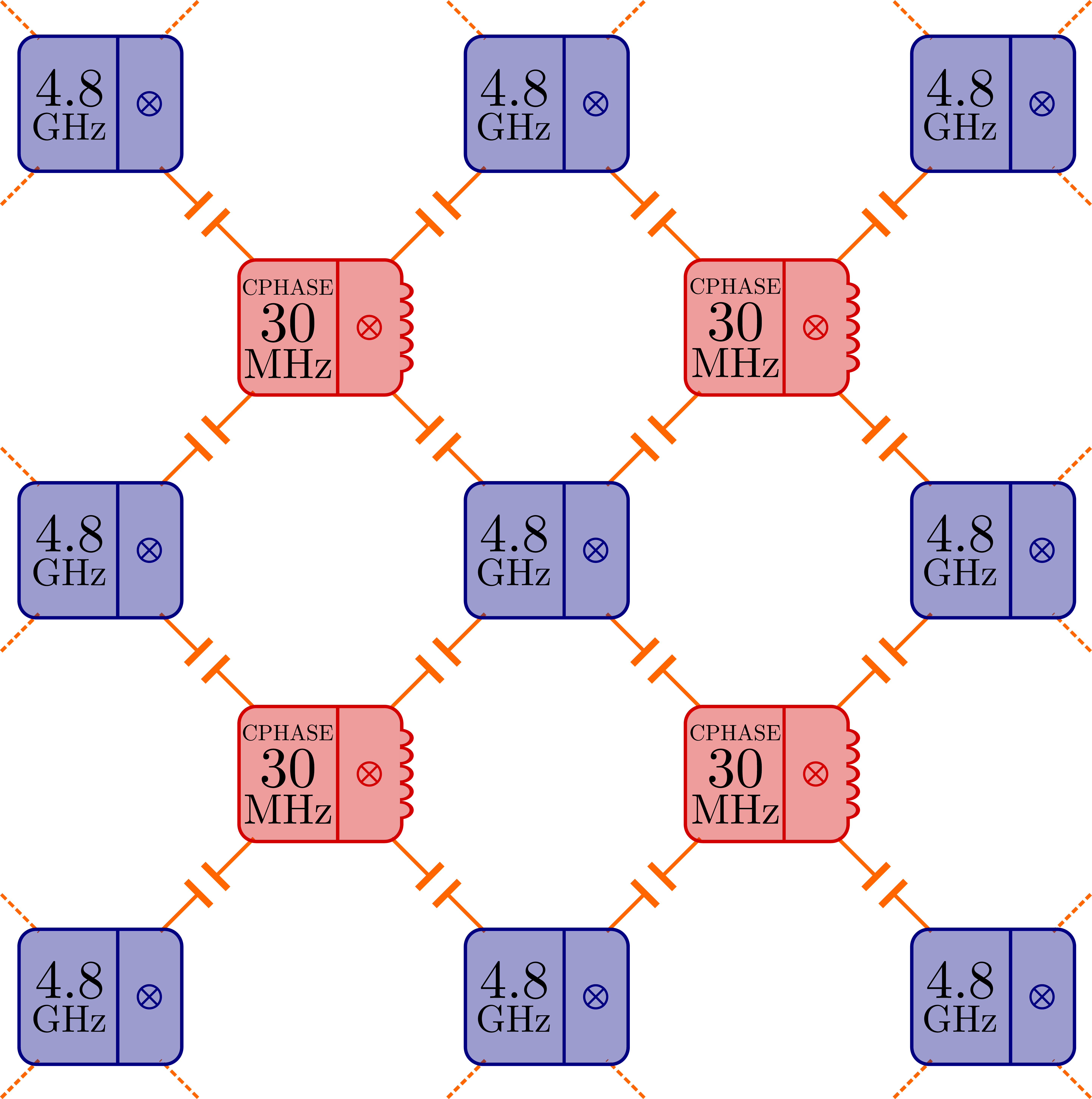}
\caption{Surface code architecture for flux-tunable transmons (blue) and low frequency fluxonia (red) based on the CPHASE gate. All transmons are chosen at the same sweet spot fundamental frequency $\omt/2\pi = 4.8 \, \mathrm{GHz}$ and anharmonicity $\anharm/2\pi = -0.3 \, \mathrm{GHz}$, while the fluxonia and the $\jc$ parameter are taken as in parameter set CPHASE in Table \ref{tab:par_set}. }
\label{fig:arch_cphase}
\end{figure}

In Fig.~\ref{fig:arch_cphase} we show a fundamental unit cell for a transmon-fluxonium chip based on the CPHASE gate described in  Sec.~\ref{sec:cphase}. In this case we require frequency-tunable transmons, since we need the ability to selectively tune the transmons to a frequency that matches approximately that of the $\ket{3}-\ket{4}$ transition of the fluxonium. This is an inevitable consequence of the fact that the gate relies on a single resonance. The transmons can all be placed approximately at the same frequency taken to be $4.8 \, \mathrm{GHz}$ in the example, while the fluxonium has target parameters equal to parameter set CPHASE in Table \ref{tab:par_set}. Thus, in order to activate the gate, we first apply a static flux to the desired transmon, in order to bring its frequency close to $\omf{34}/2 \pi=4.3 \, \mathrm{GHz}$. Then we apply a microwave tone to the fluxonium as detailed in Sec.~\ref{sec:cphase}. Typical gate times depend on the conditional phase and are estimated between $50$ and $150 \, \mathrm{ns}$ (not including the time to flux-tune the qubit back), with static $ZZ$ of $60 \, \mathrm{kHz}$ \footnote{By increasing the coupling, the gate times can be further reduced, at the price of a higher static $ZZ$ interaction.}. Thus after the gate is completed, the transmon is brought back to the sweet spot at its normal frequency. In an all fluxonium multi-qubit chip where the CPHASE gate is implemented using the scheme of Ref.~\cite{ficheux2021}, one would need to flux the fluxonia away from their flux sweet spots, i.e., the double-well configuration, in order to implement the gate selectively. This could lead to complications because additional fluxonium transitions are activated when the system is moved out of the double-well configuration. In contrast, in our scheme the fluxonia always remain at their flux sweet spots.

We remark that our procedure to selectively activate the CPHASE gate is different than the scheme used for the flux-activated CPHASE gate between two transmons \cite{rol2019} in multi-qubit architectures \cite{versluis2017}. There the flux pulse is used to implement the gate, while in addition, a neighbor qubit is parked at a different frequency to avoid a frequency collision during the gate operation. In our case, the qubits that are not involved in the gate are left untouched and the gate is activated by the microwave pulse on the fluxonium. A disadvantage is that moving the transmon away from a flux sweet spot triggers $1/f$ flux noise during the gate operation, which impacts leakage and the gate fidelity. This problem can be mitigated using Net-Zero flux pulses \cite{rol2019}. The fluxonium instead always remains at its sweet spot. Another disadvantage is that while a transmon and a fluxonium are interacting via the CPHASE gate, the other transmons coupled to the fluxonium need to remain at their sweet spots to avoid cross-driving. Therefore, these transmons cannot simultaneously interact with other fluxonia during the activation of the CPHASE gate, limiting the number of gates that can be executed in parallel in this architecture.

\subsection{Frequency collisions and chip yield study}

The transmon-fluxonium architecture based on the CR gate that we propose employs only fixed-frequency qubits. Thus, the impact of frequency crowding and the possible frequency collisions on the chip yield is an important consideration when describing the scalability of the system. In contrast, transmons are flux-tunable in the architecture using the CPHASE gate and we do not expect frequency collisions to be an issue in that case. Therefore we focus on exploring the problem of frequency crowding only in the fixed-frequency architecture. 
In this section, we discuss the fluctuations in the parameters of both fluxonia and transmons that we expect due to fabrication imprecision. 

We outline a set of frequency collisions that we expect to degrade the device performance. 
To identify the dominant collisions we numerically explore the CR gate performance as a function of the target transmon frequencies $\omt$ and extract the regions of increased leakage or crosstalk. We associate each region to a specific transition that is driven and that corresponds to a collision condition. We then define a window around each collision, inside of which the increase in crosstalk or leakage is expected to significantly degrade the gate fidelity. We perform a detailed scan in the vicinity of each collision to extract the corresponding bound by requiring that the resulting error is below a given threshold, which we specify below. For collisions involving a spectator qubit we simulate the gate using the full three-qubit system instead.
Given these bounds, we simulate the zero-collision yield over a range of variation in the tunnel junction resistance, which determines the qubit frequencies, for surface code lattices up to distance 7 and different drive amplitudes.

The transition frequencies of the transmon are determined by $\ect$ and $\ejt$, while for the fluxonium the frequencies are a function of $\ecf$, $\elf$, and $\ejf$. We do not assume any variation in the targeted $\ect$ or $\ecf$ as the shunting capacitance can be consistently reproduced~\cite{kreikebaum2020,Hertzberg2021,nguyen2022}. On the other hand, it is hard to fabricate the Josephson junction reliably, thus leading to large fluctuations in the critical current~$I_{c}$, which can be related to the tunnel barrier resistance at room temperature $R$ via the Ambegaokar-Baratoff formula $I_{c} =\frac{\pi \Delta}{2eR}$, where $\Delta$ is the superconducting energy gap. Since $R$ is readily measurable experimentally, we define the variation due to the fabrication of the Josephson junction in terms of the standard deviation $\sigma_{R}$ of the resistance. Given that the Josephson energy can be expressed as $E_{J} = \frac{\hbar I_{c}}{2e}$ we expect a variation in $\ejf$ of $\frac{\sigma_{\ejf}}{\ejf} \approx \frac{\sigma_{R}}{R}$ for the fluxonium qubit. For $\elf$, we consider the superinductance to be realized via a Josephson-junction array consisting of approximately $N\approx 100$ junctions. Therefore, independent fluctuations in the Josephson energy of each junction would lead to a variation in the inductive energy of $\frac{\sigma_{\elf}}{\elf} = \frac{\sigma_{R}}{\sqrt{N}R}$, using the approximation that the effective inductance due to the array is $L_{\rm eff}=N L_J$ \cite{ManucharyanPhd}. When the fluctuations are too large, this simple approximation may break down and one expects spectral shifts due to the coupling of the fluxonium mode with other modes of the array~\cite{FHK:symmetry}.

The transmon frequency $\omt$ is approximately given by $\omt \approx \sqrt{8\ect \ejt}/\hbar$ and we thus expect a deviation in the transmon frequency of $\frac{\sigma_{\omt}}{\omt} = \frac{\sigma_{R}}{2R}$. Given that $\ect \approx \hbar |\anharm|$ we do not expect any variation in $\anharm$. A resistance variation as low as $\sigma_{R}/R \approx 2\%$ has been previously reported~\cite{kreikebaum2020} and a variation of $\sigma_{R}/R \approx 0.5\%$ has been obtained after the use of laser annealing~\cite{Hertzberg2021}. 

\begin{table*}[ht]
  \centering
  \renewcommand{\arraystretch}{1.8}
  \begin{tabular}{| c | c | c | c | c |}
  \hline
  \multirow{2}{1.5cm}{\centering Type} & \multirow{2}{6.5cm}{\centering Frequency Collision} & \multicolumn{3}{c|}{Bounds}\\
  \cline{3-5} & & $\epsd/2\pi = 100 \, \mathrm{MHz}$ & $\epsd/2\pi = 300 \, \mathrm{MHz}$ & $\epsd/2\pi = 500 \, \mathrm{MHz}$\\
  \hline
  1 & $\omt = \omf{12}$ or $\omt = \omf{03}$ & $\pm 100 \, \mathrm{MHz}$ & $\pm 100 \, \mathrm{MHz}$ & $\pm 100 \, \mathrm{MHz}$\\
  \hline
  2 & $\omt < \omf{12}$ or $\omt > \omf{03}$ & -- & -- & --\\
  \hline
  3 & $2\omt = \omf{04}$ & $\pm 15 \, \mathrm{MHz}$ & $\pm 40 \, \mathrm{MHz}$ & $\pm 60 \, \mathrm{MHz}$\\
  \hline
  4 & $2\omt = \omf{15}$ & $\pm 5 \, \mathrm{MHz}$ & \multirow{2}{*}{$\pm 40 \, \mathrm{MHz}$} & \multirow{2}{*}{$\pm 50 \, \mathrm{MHz}$}\\
  \cline{1-3}
  5 & $2\omt = \omt + \anharm + \omf{03}$ & $\pm 9 \, \mathrm{MHz}$ &  & \\
  \hline
  6 & $3\omt = \omf{05}$ & -- & $\pm 17 \, \mathrm{MHz}$ & $\pm 35 \, \mathrm{MHz}$\\
  \hline
  7 & $\omt = \oms$ & $\pm 5 \, \mathrm{MHz}$ & $\pm 15 \, \mathrm{MHz}$ & $\pm 15 \, \mathrm{MHz}$\\
  \hline
  8 & $\omt = \oms + \delta_{s}$ & $\pm 7 \, \mathrm{MHz}$ & $\pm 20 \, \mathrm{MHz}$ & $\pm 20 \, \mathrm{MHz}$\\
  \hline
  9 & $\omt + \oms = \omf{04}$ & $\pm 10 \, \mathrm{MHz}$ & $\pm 25 \, \mathrm{MHz}$ & $\pm 50 \, \mathrm{MHz}$\\
  \hline
  \end{tabular}
  \caption{Frequency collisions and chosen bounds (on $\omega/2\pi$) on forbidden windows around the collision at three different drive amplitudes $\epsd$ for the fixed-frequency transmon-fluxonium architecture employing the CR gate. The bounds on the forbidden windows for each collision are estimated from numerical simulations with the exception of collision type 7, which is taken to be a similar ratio relative to collision type 8 as the ratio between collisions types 5 and 6 reported in~\cite{Hertzberg2021}. Collision types 1-6 involve a fluxonium acting as the control qubit and a transmon at a frequency $\omt$ acting as the target qubit. Collision types 7-9 further consider a spectator transmon at a frequency $\oms$ that is coupled to the fluxonium. The other transmon and fluxonium parameters are given in Ref.~\cite{Hertzberg2021}.}
  \label{tab:freq_colisions}
\end{table*}

Frequency collisions generally lead to an increase in the infidelity or time-duration of the targeted two-qubit gate due to an increase in leakage and crosstalk.
Based on our (noiseless) numerical simulations, we have identified the 9 most likely frequency collisions for our architecture, listed in~\cref{tab:freq_colisions}. Each collision involves either only the control fluxonium and target transmon or it further involves a spectator transmon, coupled to the fluxonium. The impact of each collision can be understood as follows: Type 1 collisions lead to a high residual $ZZ$ coupling between the transmon and the fluxonium qubit. This has to be avoided as it gives strong $ZZ$ crosstalk when we do not want to couple the qubits. We define the bound around this collision such that $\xi_{ZZ}/2\pi \leq 1 \, \mathrm{MHz}$. Collision type 2 defines conditions leading to a relatively long CR gate time $t_{\mathrm{gate}} \gg 100 \, \mathrm{ns}$, as it breaks the condition in Eq.~\eqref{eq:transmon-freq-condition}.
Type 3 (resp. type 4) collisions are two-photon transitions due to the drive at frequency $\omega_d=\omega_t$ which lead to the fluxonium leaking from $\ket{0}$ to $\ket{4}$ (resp. $\ket{1}$ to $\ket{5}$). Collision type 5 represents the drive causing the transmon to leak from $\ket{1}$ to $\ket{2}$ while the fluxonium leaks from $\ket{0}$ to $\ket{3}$. Collision type 6 leads to the fluxonium leaking from $\ket{0}$ to $\ket{5}$ via a three-photon transition due to the drive at frequency $\omega_t$. To bound each of these collisions, we require that the average leakage $L_1$ to satisfy $L_1 \leq 10^{-3}$. The next collisions consider the impact of driving a CR gate between a fluxonium and a target transmon on a spectator transmon at frequency $\oms$. Collisions type 7 leads to the drive of the CR gate addressing the neighboring spectator transmon, while type 8 collisions instead cause the spectator transmon to leak from $\ket{1}$ to $\ket{2}$. Collision type 9 describes a transition involving both the target and spectator transmon and leading to the fluxonium leaking from $\ket{0}$ to $\ket{4}$.
To extract the bounds around collision type 7, we prepare the spectator transmon in $\ket{0}$ and the control and target qubits in the  state $P_c/2$ and require that the population in $\ket{1}$ of the spectator qubit after the gate to be below $10^{-3}$.
Finally, to bound collisions type 8 and 9, we require again that the resulting average leakage from the computational states is sufficiently small, that is, $L_1 \leq 10^{-3}$. 
We note that collision types 4 and 5 are separated by only $\approx 50 \, \mathrm{MHz}$ for the targeted fluxonium parameters. A sufficiently strong drive ($\epsd/2\pi \geq 300 \, \mathrm{MHz}$) leads to a (drive-strength dependent) detuning of the qubit frequencies, which can lead to the frequencies of these two collisions to shift closer to each other. In such a case, we instead place a single bound around collision type 4, which also includes collision type 5, and ensures that the resulting leakage $L_1 \leq 10^{-3}$ outside of this window.

\begin{figure}
\centering
\includegraphics[width=0.4 \textwidth]{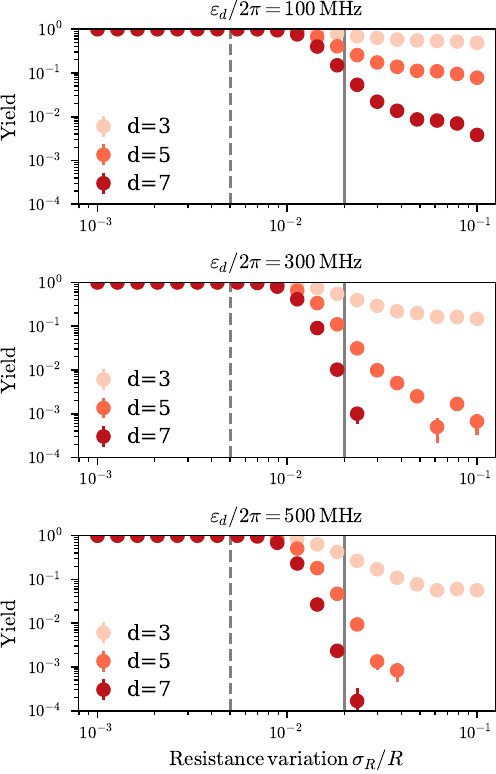}
\caption{Zero-collision yield as a function of the tunnel barrier resistance dispersion for a fixed-frequency transmon-fluxonium surface code lattice of distance $d=3, 5, 7$ (light, medium and dark red respectively) and for drive amplitudes $\epsd/2\pi = 100\, \mathrm{MHz}$ (top), $\epsd/2\pi = 300\, \mathrm{MHz}$ (middle) and $\epsd/2\pi = 500\, \mathrm{MHz}$ (bottom). The solid-gray line shows the state-of-the-art resistance variation measured, while the dashed-gray line shows the achieved variation following laser annealing. The yield is extracted over $6000$ resamples of the lattice parameters.}
\label{fig:freq_cols}
\end{figure}

To simulate the expected zero-collision yield, we consider a transmon-fluxonium surface code lattice of distance $d$ and take the fluxonia to be the ancilla qubits and the transmons to be the data qubits. We sample a $\omt$ for each transmon and $\elf$ and $\ejf$ for each fluxonium drawing from a Gaussian distribution characterized by a standard deviation determined by $\sigma_{R}$ (as described above) and centered around the targeted parameter value. We then evaluate the transition frequencies of each fluxonium via numerical diagonalization and check if any collisions have occurred across the lattice. We perform $6000$ repetitions of this process for each lattice and drive amplitude. In Fig.~\ref{fig:freq_cols} we show the results for lattices of distance $d=3, 5, 7$ and for drive amplitudes $\epsd/2\pi = 100, 300, 500\, \mathrm{MHz}$ (with the corresponding bounds given in~\cref{tab:freq_colisions}) as a function of $\sigma_{R}/R$. We observe that for a resistance variation of $\sigma_{R}/R = 2\%$ we expect all lattices up to $d = 7$ to be producible with a yield $\gtrapprox 10\%$ when $\epsd/2\pi = 100 \, \mathrm{MHz}$. When the drive amplitude is increased to $\epsd/2\pi \geq 300 \, \mathrm{MHz}$ the $d=7$ yield lattice drops to $\gtrapprox 1\%$. In the case of a strong drive of $\epsd/2\pi = 500 \, \mathrm{MHz}$, the yield for the $d=5$ and $d=7$ lattices drops to be above $1\%$ and $0.1\%$ respectively. If we instead consider the resistance variation achieved following laser annealing, we see that any lattice up to $d=7$ can be fabricated with a yield $>99\%$ for each drive amplitude considered here. In Ref.~\cite{Hertzberg2021} a transmon-transmon architecture utilizing a CR gate of duration $200-400\, \mathrm{ns}$ is explored. At the same resistance variation of~$\sigma_{R}/R \approx 0.5\%$, there is no transmon-transmon surface code lattice achieving a yield $>10\%$, while for a heavy-hexagon code lattice, only code distances $d=3, 5$ lead to a yield $>10\%$. The high yield demonstrates that the frequency crowding issue is greatly mitigated in a transmon-fluxonium architecture by the large detuning between the qubit frequencies. In~\cref{app:num_cols} we show the average number of collisions observed for a $d=3$ lattice at $\epsd/2\pi = 300\,\mathrm{MHz}$, demonstrating that the loss in yield is dominated by frequency collisions involving next-nearest-neighboring transmons. We expect a more optimal assignment of the transmon frequencies to further increase the zero-collision yield. Furthermore, a factor of 2 improvement in the resistance variation achieved from fabrication would enable a $\gtrapprox 10\%$ yield on transmon-fluxonium lattices up to distance 7, mitigating the need for post-fabrication adjustments.

\section{Conclusions}
\label{sec:conclusions}
In this paper we have studied two-qubit gates between transmons and fluxonia to be used in a multi-qubit architecture. Despite the typical large fundamental frequency difference between transmons and fluxonia, two-qubit gates are still possible thanks to the direct or indirect use of the higher levels of the fluxonium. We have analyzed two different microwave-activated gates: the CR gate and the CPHASE gate. The CR gate is suited for medium frequency fluxonia and, compared to its transmon-transmon counterpart, it can be implemented over a wider range of transmon frequencies, mitigating the frequency crowding problem. For low frequency fluxonia, the CR effect decreases and therefore we have studied a different scheme that implements a CPHASE gate using the third level of the fluxonium. While this gate is more prone to leakage, one can get arbitrary conditional phases with gate times around $100$ to $200 \, \mathrm{ns}$ and have small residual $ZZ$ coupling. We have also provided some architectural considerations for a surface-code-like architecture where each qubit is coupled to up to four neighbors. In case the architecture is based on the CR gate, it can be fully microwave-activated, while some flux control on the transmons is needed for the CPHASE case. We have shown that the fixed-frequency architecture based on the CR gate between transmons and fluxonia greatly mitigates the problem of frequency crowding. We show that a tunnel barrier resistance variation achieved by laser-annealing enables a yield of near unity for surface codes up to distance 7 and possibly higher: this is a yield which is considerably higher as compared to fixed-frequency transmon-transmon architectures using the CR gate. It would be interesting to also make a multi-qubit chip yield comparison between our transmon-fluxonium architecture and a fluxonium-fluxonium architecture as in~\cite{nguyen2022}.

\begin{acknowledgments} 
A. C. acknowledges funding from the Deutsche Forschungsgemeinschaft (DFG, German Research Foundation) under Germany's Excellence Strategy – Cluster of Excellence Matter and Light for Quantum Computing (ML4Q) EXC 2004/1 – 390534769 and from the German Federal Ministry of Education and Research in the funding program ``quantum technologies – from basic research to market" (contract number 13N15585). B.V. and B. M. T. are supported by QuTech NWO funding 2020-2024 – Part 1 “Fundamental Research” with project number 601.QT.001-1. C.K.A. is supported by QuTech NWO funding 2020-2024 - Part 1-3. The numerical analysis was carried out using an in-house Python package for superconducting qubits available on github at \url{https://github.com/cianibegood/pysqkit.git}, that relies on the QuTiP package for the dynamical simulations \cite{qutip}. The data to obtain the figures are available at \textcolor{blue}{\doi{10.5281/zenodo.6628824}}.
\end{acknowledgments} 

\appendix

\section{Schrieffer-Wolff and CR gate analysis}
\label{app:sw}

In this Appendix we perform a perturbative analysis of the coupled transmon-fluxonium system based on the Schrieffer-Wolff transformation, following closely Ref.~\cite{BRAVYI20112793, magesan2020, juelich_le}. We first execute the analysis without the drive and obtain perturbative formulas for the frequency shifts of the levels as well as for the residual $ZZ$ coupling. 
In our analysis, we include the first $3$ levels of the transmon and the first $4$ levels of the fluxonium.  Within this subspace the Hamiltonian in Eq.~\eqref{eq:h} reads 

\begin{equation}
H = H_0 + V,
\end{equation}
with the Hamiltonians $H_0$ and $V$ which, in the bare basis, are 
\begin{multline}
\frac{H_0}{\hbar} = \omt\ketbrat{1}{1} + (2 \omt + \anharm) \ketbrat{2}{2} \\
+ \sum_{k=1}^3 \omf{k} \ketbraf{k}{k},
\end{multline}
and
\begin{multline}
\label{eq:vrwa}
V \overset{\mathrm{RWA}}{=} \jc \qzpf \bigl(\sigmat{0}{1} + \sqrt{2} \sigmat{1}{2} \bigr) \times \\
\bigl(\qfel{1}{0} \sigmaf{1}{0} + \qfel{3}{0} \sigmaf{3}{0} + \qfel{2}{1} \sigmaf{2}{1}  + \qfel{3}{2} \sigmaf{3}{2}  \bigr) + \mathrm{h.c.}
\end{multline}
Here $\ketbrat{k}{k} = \ketbra{k}{k}_0 \otimes I_{0f}$ and $\ketbraf{k}{k} = I_{0t} \otimes \ketbra{k}{k}_0$, with $I_{0t}$ and $I_{0f}$ the identity on the transmon and fluxonium, respectively. Analogously,  $\sigma_{kl}^{t} = \ketbra{k}{l}_0 \otimes I_{0f}$ and $\sigma_{kl}^{f} = I_{0t} \otimes \ketbra{k}{l}_0$, while the $\qfel{k}{l}$ are given in Eq.~\eqref{eq:qfel} in the main text. Furthermore, the $\omf{k}/2 \pi$ are the frequencies associated with the fluxonium levels, i.e., with respect to the ground state of the fluxonium, while $\omt/2 \pi$ and $\anharm /2 \pi$ the fundamental frequency and the anharmonicity of the transmon, respectively. So in this Appendix, for notational simplicity, $\omf{0}$ is set as 0 (while the explicit dependence is given in expressions in the main text and Eq.~\eqref{eq:transmon-freq-condition}). 

In Eq.~\eqref{eq:vrwa} we performed a rotating wave approximation (RWA) neglecting terms $\sigma_{kl}^{t} \sigma_{k'l'}^f$ with $k>k', l>l'$, and their Hermitian conjugate.

In order to proceed with the Schrieffer-Wolff analysis let us define the relevant projectors in the bare basis
\begin{subequations}
\begin{equation}
P_{0}= \sum_{k, l=0}^1\ketbra{kl }{kl }_0, 
\end{equation}
\begin{equation}
Q_0 = I - P_{0}, 
\end{equation}
\end{subequations}
and in the dressed basis
\begin{subequations}
\begin{equation}
P= \sum_{k, l=0}^1 \ketbra{kl}{kl}, 
\end{equation}
\begin{equation}
Q = I - P.
\end{equation}
\end{subequations}
The Schrieffer-Wolff transformation is defined as the unitary $U$ that transforms the projectors in the bare basis to those in the dressed basis:
\begin{equation}
U^{\dagger} P_0 U = P, \quad U^{\dagger} Q_0 U = Q.
\end{equation}
The unitary $U$ exists and is unique if and only if $\lVert P - P_0 \rVert < 1$ and in this case it is given by \cite{BRAVYI20112793}
\begin{equation}
\label{eq:u_sw_exact}
U = \sqrt{(I - 2 P_0)(I-2P)}.
\end{equation}
In addition, $U$ can be written as $U = \exp(S)$ with $S$ being anti-Hermitian, $S = -S^{\dagger}$, and block-off-diagonal with respect to $P_0$, i.e., $P_0 S P_0 = (I - P_0) S (I - P_0)=0$. The goal of the Schrieffer-Wolff transformation is to obtain an effective Hamiltonian $H_{\mathrm{eff}}$ that has the same spectrum as $P H P $, that is, the projection of $H$ onto the subspace associated with $P$, i.e., the computational subspace in our case. The effective Hamiltonian is given by 
\begin{equation}
\label{eq:h_sw_eff}
H_{\mathrm{eff}} = P_0 U H U^{\dagger} P_0.
\end{equation}   
Eq.~\eqref{eq:u_sw_exact} provides a numerical method to obtain the Schrieffer-Wolff unitary and thus also the effective Hamiltonian \cite{Consani_2020}. However, the standard use of the Schrieffer-Wolff is to find an analytical, perturbative expansion of the effective Hamiltonian. The norm of the coupling operator $\lVert V \rVert$ quantifies the strength of the coupling, and thus plays the role of the coupling parameter. Following Ref. \cite{BRAVYI20112793}, the second-order expansion of the effective Hamiltonian is given by
\begin{equation}
H_{\mathrm{eff}}^{(2)} = P_0 (H_0 + V) P_0 + \frac{1}{2} P_0 [S_1, V_{\mathrm{od}}] P_0,
\end{equation}   
where $V_{\mathrm{od}} = P_0 V Q_0 + Q_0 V P_0$ is the off-diagonal part of $V$. In our case $V$ is fully off-diagonal, i.e. $V_{od}=V$, and the anti-Hermitian operator $S_1$ reads
\begin{multline}
S_1 = \frac{\jc \qzpf \qfel{3}{0}}{\hbar(\omf{3} - \omt)} \sigmat{0}{1} \sigmaf{3}{0} + \frac{\jc \qzpf \qfel{2}{1}}{\hbar[(\omf{2} - \omf{1}) - \omt]} \sigmat{0}{1} \sigmaf{2}{1} \\ + \frac{\sqrt{2} \jc \qzpf \qfel{1}{0}}{\hbar[\omf{1} - ( \omt + \anharm )]} \sigmat{1}{2} \sigmaf{1}{0} - \mathrm{h.c.}.
\label{eq:defS1}
\end{multline}
This gives the effective Hamiltonian
\begin{multline}
\label{eq:heff_sw}
\frac{H_{\mathrm{eff}}^{(2)}}{\hbar} = \omt\ketbrat{1}{1} + \omf{1} \ketbraf{1}{1} \\
+ \jc \qzpf \qfel{1}{0} \bigr[\sigmat{0}{1} \sigmaf{1}{0} + \mathrm{h.c.} \bigl] \\ 
+ \zeta_{10} \ketbra{10}{10}_0 +\zeta_{11} \ketbra{11}{11}_0,
\end{multline}

where we define the coefficients
\begin{subequations}
\begin{equation}
\hbar \zeta_{10} = -\frac{\jc^2 \qzpf^2 \qfel{3}{0}^2}{\omf{3} - \omt},
\end{equation}
\begin{equation}
\hbar \zeta_{11} = \jc^2 \qzpf^2 \biggl[\frac{2 \qfel{1}{0}^2}{\omf{1} - (\omt + \anharm)} - \frac{\qfel{2}{1}^2}{(\omf{2} - \omf{1}) - \omt} \biggr],
\label{eq:z11}
\end{equation}
\end{subequations}
We note that the first contribution in $\zeta_{11}$ is due to the $\ket{11}_0-\ket{20}_0$ transition. It is quite small as $q_{f,10}$ is small (see Fig.~\ref{fig:flx_mat_elem}), and the levels are fairly off-resonant (see Fig.~\ref{fig:tr_flx}).
The second term is dominant and due to the (more resonant) $\ket{11}_0-\ket{02}_0$ transition with larger $q_{f,21}$ in Fig.~\ref{fig:tr_flx}. The sign of the contributions is opposite as $\ket{20}_0$ is higher than $\ket{11}_0$ while $\ket{02}_0$ is lower. The coefficient $\zeta_{10}$ is due to the $\ket{10}_0-\ket{03}_0$
transition.

From Eq.~\eqref{eq:heff_sw} we see that we still have the exchange coupling term proportional to $\jc$ that couples levels $\ket{10}_0$ and $\ket{01}_0$. This could be further removed with a second Schrieffer-Wolff transformation that would give two effective Hamiltonians: one for the subspace $\{\ket{00}_0, \ket{01}_0\}$ and one for the subspace $\{\ket{10}_0, \ket{11}_0 \}$. However, we remark that the second-order corrections due to this flip-flop term are very small in our typical setup because $|\qfel{1}{0}| \ll 1$ and, in addition, the transmon and fluxonium frequency always differ by at least $3.5 \, \mathrm{GHz}$. Thus, we can simply neglect the effect of this term. Within this approximation, the $ZZ$ coupling in second-order Schrieffer-Wolff is given by
\begin{multline}
\label{eq:zzsw}
\hbar \xi_{ZZ}^{(\mathrm{SW})}= \bra{11}H_{\mathrm{eff}}^{(2)} \ket{11}_0 + \bra{00}H_{\mathrm{eff}}^{(2)} \ket{00}_0 - \\ 
\bra{01}H_{\mathrm{eff}}^{(2)} \ket{01}_0 - \bra{10}H_{\mathrm{eff}}^{(2)} \ket{10}_0  = \hbar(\zeta_{11} - \zeta_{10}).
\end{multline}
Thus the $ZZ$ coupling is enhanced in strength when the signs of $\zeta_{11}$ and $\zeta_{10}$ are opposite.
Neglecting the first term in Eq.~\eqref{eq:z11}, this is achieved when the transmon frequency is chosen as
\begin{align}
 \omega_{f,2}-\omega_{f,1}   < \omega_t < \omega_{f,3}-\omega_{f,0}. 
 \label{eq:transmon-freq-condition}
\end{align}

In a multi-qubit architecture where all qubits are capacitively coupled, the SW transformation to the dressed, computational basis, will not only slightly entangle nearest-neighbor capacitively-coupled qubits but also entangle non-nearest neighbors. That is, $i S_1$ will be a 2-local many-qubit Hamiltonian with non-commuting 2-local terms each representing the nearest-neighbor qubit couplings. This implies that the computational qubits, on which we also apply single-qubit gates and which we measure, are represented by two-level subspaces which are partially multi-qubit entangled. As a consequence, a drive on one qubit in the bare basis, will be transferred not only to its nearest-neighbor qubits, but also, more weakly, to non-nearest neighbor qubits.

\subsection{Drive and CR coefficient}

In the presence of a drive on the fluxonium the Hamiltonian gets an additional term given in Eq.~ \eqref{eq:hdrive}. We can use the previous analysis to get an expression for the CR effect by simply applying the Schrieffer-Wolff transformation to the drive Hamiltonian \cite{magesan2020, juelich_le, joelthesis} so we see the effect of the drive in the dressed, computational, basis.

In what follows we assume, for simplicity, that the envelope function is a constant, i.e., $g(t)=1$, and set $\phased = \pi/2$, but we will comment on what happens when we change $\phased$.

After obtaining the effective Hamiltonian in Eq.~\eqref{eq:heff_sw}, we switch to a rotating (or interacting) frame at the drive frequency for both qubits defined by the reference Hamiltonian $H_{\mathrm{ref}}/\hbar = \omd(\ketbrat{1}{1} + \ketbraf{1}{1})$, for the purpose of analysis. In general, for a Hamiltonian $H$, moving to rotating frame set by $U_{\rm ref}=e^{-i H_{\rm ref}t/\hbar}$ results in the dynamics being given by a Hamiltonian $\tilde{H}$ given by 
\begin{align}
    \tilde{H}(t) = U_{\rm ref}^{\dagger} H U_{\rm ref} +i \left(\frac{d}{dt} U_{\rm ref}^{\dagger}\right) U_{\rm ref}= \notag \\
    e^{i H_{\mathrm{ref}} t/\hbar} H e^{-i H_{\mathrm{ref}} t/\hbar}- H_{\mathrm{ref}}.
\end{align}

We then approximate and calculate
\begin{multline}
\frac{\tilde{H}_{\mathrm{drive, eff}}(t)}{\hbar} = \\
e^{i H_{\mathrm{ref}}   t/\hbar} P_0 e^{S_1} \frac{H_{\mathrm{drive}}(t)}{\hbar} e^{-S_1} P_0 e^{-i H_{\mathrm{ref}/\hbar}   t} 
\approx \\
e^{i H_{\mathrm{ref}/\hbar}   t} P_0 \frac{\left( H_{\mathrm{drive}}(t) +[S_1,H_{\rm drive}]\right)}{\hbar} P_0 e^{-i H_{\mathrm{ref}/\hbar}   t} 
\overset{\mathrm{RWA}}{\approx} \\
\mu_{X_{f}} X_f + \mu_{\mathrm{CR}} X_t Z_f + \mu_{X_t} X_t.
 \label{eq:drive-eff-rot}
\end{multline}
Here the first term is simply due to $H_{\rm drive}(t)$ and $[S_1,H_{\rm drive}]$ gives the other two terms, i.e. the (fluxonium-controlled) rotation on the transmon qubit. Here $X_f$($X_t$) is the Pauli $X$ operator acting on the fluxonium (transmon), and $Z_f$ the Pauli $Z$ acting on the fluxonium. 

The coefficients within this approximation read
\begin{subequations}
\begin{equation}
\hbar \mu_{X_f} = \frac{1}{2} \qfel{1}{0}\epsd
\end{equation}
and
\begin{equation}
\hbar \mu_{X_t} = \frac{\jc \qzpf}{4} \biggl[\frac{\qfel{3}{0}^2}{\omt - \omf{3}} + \frac{\qfel{2}{1}^2}{\omt -(\omf{2} - \omf{1})}  \biggr] \epsd,
\end{equation}
\end{subequations}
while the CR coefficient is 
\begin{equation}
\label{eq:mu_cr_sw}
\hbar \mu_{\mathrm{CR}} = \frac{\jc \qzpf}{4} \biggl[\frac{\qfel{3}{0}^2}{\omt - \omf{3}} - \frac{\qfel{2}{1}^2}{\omt -(\omf{2} - \omf{1})} \biggr] \epsd.
\end{equation}

As for the strength of the CR coefficient $\mu_{\rm CR}$, we can observe the following. Similar as for the $ZZ$ coupling, the largest coefficient $\mu_{\rm CR}$ is obtained when the two terms in Eq.~\eqref{eq:mu_cr_sw} add constructively, i.e. we choose the transmon frequency according to Eq.~\eqref{eq:transmon-freq-condition}.
Naturally, the more entangling the Schrieffer-Wolff unitary $e^{S_1}$ to the dressed basis is, the more the drive on the bare transmon qubit becomes transferred to a coupling term and this entangling power of $e^{S_1}$ with $S_1$ in Eq.~\eqref{eq:defS1} is proportional to $J_C$. Additionally, our perturbative formula Eq.~\eqref{eq:mu_cr_sw} predicts a linear increase of $\mu_{\mathrm{CR}}$ with the drive amplitude $\epsd$. As for the transmon-transmon case, a more refined analysis that includes the drive in the perturbation would predict a saturation of the cross-resonance coefficient with the drive amplitude \cite{magesan2020}. By adapting this analysis to our case, we verified that for the parameters used in this manuscript the linear approximation Eq.~\eqref{eq:mu_cr_sw} is quite accurate and reproduces the ``exact" result with error below $5 \%$ for all the considered drive amplitudes.

Notice that if $\mu_{\mathrm{CR}}$ is negative for $\phased= \pi/2$, we can always change its sign by taking $\phased = 3\pi/2$ instead in Eq.~\eqref{eq:hdrive}. Thus, $\mu_{\mathrm{CR}}$ can always be assumed to be positive. 
By choosing a different phase $\theta_d$ of the drive, say $\theta_d=0$, one can go through the math behind Eq.~\eqref{eq:drive-eff-rot} and observe that Pauli $X_t \rightarrow Y_t$ and $X_f \rightarrow Y_f$ as one may expect.

In the rotating frame set by $H_{\rm ref}$, Eq.~\eqref{eq:heff_sw} equals
\begin{align}
    \frac{\tilde{H}_{\rm eff}^{(2)}}{2 \hbar} \approx (\omega_t-\omega_d) \ket{1}\bra{1}_t+(\omega_{f,1}-\omega_d)\ket{1}\bra{1}_f+\notag \\
    \zeta_{10}\ket{10}\bra{10}+\zeta_{11}\ket{11}\bra{11},
\end{align}
neglecting the flip-flop coupling in the computational subspace. Hence, we see that if $\omd$ is chosen as $\omt$, one drives the fluxonium at the frequency of the transmon qubit, $\tilde{H}_{\rm eff}^{(2)}$ contains no single-qubit $Z_t$ and the effect of the $\mu_{\rm CR} X_t Z_f$ and $\mu_{X_t} X_t$ terms in Eq.~\eqref{eq:drive-eff-rot} is maximal. If instead the drive frequency and the transmon frequency are sufficiently different, the (fluxonium-controlled) $X_t$-rotation is very small as compared to $Z_t$, and hence would induce at most some renormalization of the transmon frequency and the $ZZ$-coupling.

In order to completely understand the dynamics, we can re-evaluate this Hamiltonian in the standard computational rotating frame where the precession of each (dressed) qubit at its eigenfrequency is undone. For the transmon qubit, $H_{\rm ref}$ already selects this frame (as $\omega_d=\omega_t$), but for the fluxonium qubit we can undo the first rotation frame and use the first computational one for the fluxonium by picking a new $H_{\rm ref}'/\hbar=(\omega_{1,f}-\omega_d) \ket{1}
\bra{1}_f$.  This ensures that $\tilde{H}_{\rm eff}^{(2)}$ has no more single-qubit $Z_t$ or $Z_f$ terms, and note it does not affect the relevant $X_t Z_f$ term. At the same time it introduces a time-dependence $e^{\pm 2i  (\omega_{1,f}-\omega_d) t}$ in the single-qubit term $\mu_{X_f} X_f$ in Eq.~\eqref{eq:drive-eff-rot}. Both due to the smallness of $q_{f,10}$ as well as the difference in frequency between the transmon and the fluxonium (Fig.~\ref{fig:flx_mat_elem}), the effect of this off-resonant term is thus very small.

\begin{figure}
\vspace{0.5 cm}
\centering
\includegraphics[width=0.4 \textwidth]{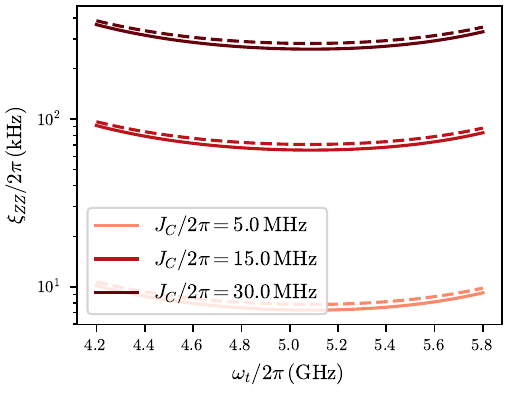}
\caption{Residual $ZZ$ coupling as a function of the transmon frequency at zero drive strength. The solid lines represent the exact numerical value, while the dashed line shows the result obtained from the second-order Schrieffer-Wolff transformation in Eq.~\eqref{eq:zzsw}. We use the fluxonium parameters of parameter set CR in Table \ref{tab:par_set}.}
\label{fig:sw_comp}
\end{figure}

In Fig.~\ref{fig:sw_comp} we compare the results obtained with the perturbative Schrieffer-Wolff analysis and the exact numerical values for the $ZZ$ coupling in Eq.~\eqref{eq:zzsw}. We see good agreements between the numerics and the results from the perturbative Schrieffer-Wolff analysis. We notice that the $ZZ$ coupling is relatively constant over a wide range of frequencies.  This is due to the fact that the transmon frequency is always in between the transition frequencies of the $\ket{1}-\ket{2}$ and the $\ket{0}-\ket{3}$ transition of the fluxonium, as in Eq.~\eqref{eq:transmon-freq-condition}. 

We conclude this Appendix by commenting on the role of the coupling parameter $\jc$ on the gate implementation. The unwanted $ZZ$ coupling coefficient depends quadratically on $\jc$, while the cross-resonance coefficient scales linearly with $\jc$. This suggests that smaller $\jc$ should decrease the error associated with the coherent $ZZ$ interaction. However, smaller $\jc$ also means longer gate times, which we would like to limit in order to have small errors from decoherence processes. We remark that the transmon-transmon implementation of the cross-resonance gate faces exactly the same trade-off. 

\section{CPHASE gate analysis}
\label{app:cphase}

In this Appendix we provide a simplified analysis of the CPHASE gate, restricting ourselves to the computational subspace plus the higher (dressed) levels $\ket{03}$ and $\ket{13}$ to understand the idea behind the gate. We will work in the dressed computational basis $\ket{kl}$ which can be obtained from the bare basis $\ket{kl}_0$ by a Schrieffer-Wolff transformation which is discussed in Appendix \ref{app:sw}. This transformation gives a $ZZ$ coupling between the qubits (and a very off-resonant flip-flop interaction which we neglect here) which is used for the CPHASE gate.

We assume that the drive $H_{\rm drive}(t)$ in Eq.~\eqref{eq:hdrive} has $g(t)=1$ and the phase is chosen as $\theta_d=\pi/2$. We solely focus on the drive enacting the transitions $\ket{00} \leftrightarrow \ket{03}$ and $\ket{10} \leftrightarrow \ket{13}$ where $\ket{kl}$ are dressed energy levels, so we write
\begin{align}
    \frac{H_{\rm drive}(t)}{\hbar} \approx \frac{\epsd q_{f,00-03}}{2}(\ket{00}\bra{03}(-e^{i \omega_d t}+e^{-i\omega_d t})+{\rm h.c.}) \notag \\ +\frac{\epsd q_{f,10-13}}{2}(\ket{10}\bra{13}(-e^{i \omega_d t}+e^{-i\omega_d t})+{\rm h.c.}),
\end{align}
with $\qfel{kl}{-mn}=|\bra{kl}\qf \ket{mn}|$.
The rest of the Hamiltonian of the transmon-fluxonium system, restricted to this six-dimensional subspace equals
\begin{align}
    \frac{H|_6}{\hbar}=\omega_{01}P_{01}+ \omega_{11}P_{11} 
+     \omega_{00} P_{00}+ \omega_{03}P_{03} \notag \\
+    \omega_{10}P_{10}+\omega_{13}P_{13},
\end{align}
where $P_{ij}=\ket{ij}\bra{ij}$.
Due to the $ZZ$ coupling, $\Delta$ defined in Eq.~\eqref{eq:defdelta} is unequal to zero and the entangling rate of the gate is (roughly) determined by $\Delta$ since driving an uncoupled fluxonium qubit could not generate a CPHASE entangling gate. Going to a rotating (interaction) frame with reference Hamiltonian $H_{\rm ref}/\hbar=\omega_d (P_{03}+P_{13})$ (and neglecting fast-rotating terms depending on $e^{\pm 2\omega_d t}$) gives a time-independent Hamiltonian
\begin{align}
    \lefteqn{\frac{\tilde{H}}{\hbar}=\omega_{01}P_{01}+ \omega_{11}P_{11}} &
    \notag \\ & + \omega_{00}  P_{00}+(\omega_{03}-\omega_d)P_{03}-\frac{\epsd q_{f,00-03}}{2}(\ket{00}\bra{03}+{\rm h.c.})\notag \\
   &  +\omega_{10}P_{10}+(\omega_{13}-\omega_d)P_{13}-\frac{\epsd q_{f,10-13}}{2}(\ket{10}\bra{13}+{\rm h.c.}).
    \end{align}
Thus we see that one is driving Rabi oscillations in two effective qubit subsystems, namely the two-level subsystem $\ket{00}-\ket{03}$ and the two-level subsystem $\ket{10}-\ket{13}$.
For a qubit Hamiltonian $H_{\rm qubit}/\hbar=\begin{pmatrix} \alpha & \gamma/2 \\ \gamma/2 & \beta \end{pmatrix}$, one can use that 
\begin{align}
    U(t)=e^{-i H_{\rm qubit} t/\hbar} =e^{-i{\rm Tr}(H_{\rm qubit})t/(2\hbar)}e^{-i \theta \hat{n} \cdot \vec{\sigma}/2},
\end{align} 
with angle $\theta=t\sqrt{(\alpha-\beta)^2+\gamma^2}$. A full Rabi oscillation, which induces no leakage, occurs for $\theta=2\pi$, so that $e^{-i \pi \hat{n} \cdot \vec{\sigma}}=-I$, $t_{\rm gate}=\frac{2\pi}{\sqrt{(\alpha-\beta)^2+\gamma^2}}$ and $U(t_{\rm gate})=-e^{-i(\alpha+\beta)t_{\rm gate}/2}I$.
Applying this to the simultaneous Rabi oscillations in the two subspaces, we see that one needs to fulfill the condition in Eq.~\eqref{eq:cphase_cond} to get a full Rabi oscillation in both qubit subspaces.
For this $t_{\rm gate}=2\pi/\Omega$, the phases picked up by the computational states are
\begin{align}
    \phi_{00} & = \pi- (\omega_{00}+\omega_{03}-\omega_d) t_{\rm gate}/2, \notag \\
    \phi_{01} & =  - \omega_{01} t_{\rm gate} \notag \\
    \phi_{10}&= \pi- (\omega_{10}+\omega_{13}-\omega_d) t_{\rm gate}/2 \notag \\
    \phi_{11} & = -\omega_{11} t_{\rm gate}.
\end{align}
A CPHASE gate can be brought to the form in Eq.~\eqref{eq:cphase} by single-qubit $Z$ gates with $\phi=\phi_{11}-\phi_{10}-\phi_{01}+\phi_{00}
\approx \frac{t_{\rm gate}}{2}(\omega_{13}-\omega_{10}-\omega_{03}+\omega_{00})=\frac{t_{\rm gate}}{2}\Delta=\frac{\pi\Delta}{\Omega}$ where we have neglected the effect of the $ZZ$ coupling in the computational subspace. Hence $\phi \approx \pi \Delta/\Omega$. A given targeted phase $\phi$ thus leads to a targeted $\Omega$ (which sets the gate time $t_{\rm gate}$), and the targeted $\Omega$ is used to solve for a drive-frequency $\omega_d$ and a drive-power $\epsd$ which satisfies Eq.~\eqref{eq:cphase_cond}. The frequency $\omega_d$ is chosen to be close to $\omega_{03}-\omega_{00}$ and $\omega_{13}-\omega_{10}$, say midway between those transitions. The spectrum of the fluxonium and transmon qubit should be such that this choice of $\omega_d$ avoids it being close to other fluxonium transitions, such as $\ket{01} \leftrightarrow \ket{04}$.

We note that changing the phase $\theta_d$ of the drive has no effect on the gate as it simply changes the Rabi driving to be around an axis in the $XY$-plane instead of around the $X$-axis. We also note that if the transmon qubit is flux-tunable, one can vary $\Delta$ (letting it range from, say, negative to positive), and hence get a varying phase at a fixed $t_{\rm gate}$.

\section{Fidelity and leakage definitions}
\label{app:fid}
We take the definitions of gate fidelity and leakage from Ref.~\cite{wood2018}. We report them here for completeness with some notational adaptation. Let $P_c$ denote the projector onto the computational subspace encoding $n$ qubits and $d_{\mathrm{c}}=2^{n}$ be its dimension. Let $\mathcal{U}$ be the quantum operation associated with a target unitary $U$ that we want to implement within the computational subspace and that acts as the identity on the leakage subspace, i.e. $\mathcal{U}(\rho) = U \rho U^{\dagger}$ and ${\mathcal U}^{\dagger}$ is its inverse. Let $\mathcal{E}$ be the quantum operation we actually apply to the system. The average gate fidelity within the computational subspace is given by

\begin{equation}
F_{\mathrm{gate}} = \int d \psi_c \bra{\psi_c} U^{\dagger} \mathcal{E}(\ketbra{\psi_c}{\psi_c}) U \ket{\psi_c},
\end{equation}
where $d \psi_c$ denotes the Haar measure over states in the computational subspace. The process or entanglement fidelity in the computational subspace equals 
\begin{align}
F_{\mathrm{ent}} = \bra{\Psi_c} I \otimes {\cal U}^{\dagger}\circ {\cal E}(\ket{\Psi_c}\bra{\Psi_c}) \ket{\Psi_c},
\end{align} 
where $\ket{\Psi_c}=\frac{1}{\sqrt{d_c}}\sum_{i=1}^{d_c} \ket{i,i}$ is the maximally-entangled state in the computational subspace. The average leakage is defined as
\begin{equation}
L_1 = 1 - \frac{1}{d_c} \mathrm{Tr}\bigl[P_c\, \mathcal{U}^{\dagger} \circ \mathcal{E}(P_c) \bigr]. 
\end{equation}

For any trace-preserving channel ${\cal S}$ on a $d_c$-dimensional system we have the relation $F=\frac{d_c F_{\rm ent}+1}{d_c+1}$ where $F$ is the fidelity. Here, the effective channel on the computational space is ${\cal S}(\rho)=P_c [\mathcal{U}^{\dagger} \circ {\cal E}(P_c \rho P_c)] P_c$ which is not trace-preserving but trace-decreasing and ${\rm Tr}\,{\cal S}(I/d_c)=1-L_1$. Incorporating this trace-decreasing property into the standard derivation \cite{nielsen-fid} of the relation between process fidelity and gate fidelity gives 
\begin{equation}
F_{\mathrm{gate}} = \frac{d_c F_{\mathrm{ent}} + 1 - L_1}{d_c + 1}.
\label{eq:gate-process}
\end{equation}
Thus to compute the gate fidelity, one computes the entanglement or process fidelity $F_{\rm ent}$ and $L_1$. In turn, $F_{\rm ent}$ can be re-expressed, using $\ket{\Psi_c}\bra{\Psi_c}=\frac{1}{d_c}\sum_\mu P_{\mu} \otimes P_{\mu}$ with (normalized) Pauli matrices $P_{\mu}$ (${\rm Tr}\, P_{\mu} P_{\nu}=\delta_{\mu \nu}$) as 
\begin{align}
F_{\rm ent}=\frac{1}{d_c^2}\sum_{\mu=1}^{d_c^2}{\rm Tr}( U P_{\mu} U^{\dagger}{\cal E}(P_{\mu})).
\end{align}
So one evaluates ${\cal E}$ on the Pauli eigenstates of $P_{\mu}$ and takes the expectation value with the appropriate observable $U P_{\mu} U^{\dagger}$ etc. to compute $F_{\rm ent}$. $L_1$ can be computed similarly, resulting through Eq.~\eqref{eq:gate-process} in the evaluation of $F_{\rm gate}$.

\begin{table*}[t]
  \centering
\renewcommand{\arraystretch}{1.8}
  \begin{tabular}{c | P{1.7cm}| P{1.7cm}| P{1.7cm} | P{1.7cm} | P{1.7cm}}
  & \multicolumn{4}{c}{Fluxonium} \\
  \hline
    Parameter set & $T_{1}^{0 \mapsto 1} \, (\mathrm{\mu s})$ & $T_{1}^{2 \mapsto 1} \, (\mathrm{\mu s})$&  $T_{1}^{4 \mapsto 1} \, (\mathrm{\mu s}) $& $T_{1}^{3 \mapsto 2} \, (\mathrm{\mu s}) $& $T_{1}^{4 \mapsto 3} \, (\mathrm{\mu s}) $\\
 
 \hline
    CR & $510$ & $9$ & $60$ & $8$ & $4$ \\
  \hline
  CPHASE & $3976$ & $7$ & $90$ & $81$ & $4$ 
    
  \end{tabular}
  \caption{Relaxation and excitation times for other relevant fluxonium transitions for the CR and CPHASE parameter set in Table~\ref{tab:par_set}. The dielectric loss tangent and the temperature of the environment are taken as described in the caption of Table~\ref{tab:par_set}.}
  \label{tab:t1_flx}
\end{table*}

\section{Error model}
\label{app:error_model}
In the main part of the manuscript we have shown results of simulations under coherent and noisy evolutions. In this Appendix we detail our noise model. As noise source we considered only relaxation due to dielectric losses. We do not include the effect of $1/f$ flux noise in the fluxonium since it is always assumed to be biased at $\phi_{\mathrm{ext}, f}=\pi$, which is a flux sweet spot. Clearly in a multi-qubit architecture with tunable transmons that could be biased away from the flux-insensitive point this source of error would play a role, similarly to the CPHASE gate in transmon-transmon architectures \cite{rol2019}. We do not include pure dephasing mechanisms, since the model of dephasing highly depends on the experimental setup one considers. Thus, while this noise source should be included when modeling an experiment, we left it out from our analysis.  

Errors are assumed to be Markovian and modelled via a Lindblad master equation of the following form  \cite{BreuerPetruccione}
\begin{equation}
\frac{d \rho}{d t} = \frac{1}{i \hbar} [H(t), \rho] + \sum_{k} \mathcal{D}[L_k](\rho),
\end{equation}
where $\rho$ is the density matrix for the system, $H(t)$ is a general time-dependent Hamiltonian and $\mathcal{D}$ is the Lindblad dissipator
\begin{equation}
\mathcal{D}[L_k](\rho) = L_k \rho L_k^{\dagger} - \frac{1}{2}L_k^{\dagger} L_k \rho - \frac{1}{2} \rho L_k^{\dagger} L_k, 
\end{equation}
with $L_k$ the so-called jump operators. In the following we specify the form of the jump operators modeling dielectric loss.

For both transmon as well as fluxonium qubits, dielectric loss can be modelled by adding a real part to the admittance (in the frequency domain) of the shunting capacitance \cite{nguyen2019}. More precisely, this admittance is assumed to be of the following form
\begin{equation}
Y_{C}(\omega) = \frac{\omega C}{Q_{\mathrm{diel}}} + i \omega C,
\end{equation}
with $Q_{\mathrm{diel}}$ the quality factor related to the dielectric loss tangent, namely $Q_{\mathrm{diel}} = 1/\tan \delta_{\mathrm{diel}} $. The dielectric loss tangent can in turn be frequency-dependent \cite{nguyen2019} and we consider this in the case of the fluxonium, so $Q_{\mathrm{diel}} = Q_{\mathrm{diel}}(\omega)$ (see caption of Table \ref{tab:par_set} in the main text).
The following discussion applies to both transmon and fluxonium. Let $\ket{k}$ and $\ket{l}$ be a pair of (bare) energy levels with transition frequency $\omega_{kl} = (E_k - E_l)/\hbar$ with $E_{k}$ and $E_l$ the energies associated with the levels and $k > l$. The decay rate from level $\ket{k}$ to $\ket{l}$ at temperature $T=0$ reads
\begin{equation}
\gamma_{k l} = \frac{\Phi_0^2}{\hbar 2 \pi^2} \lvert \bra{k} \phi \ket{l} \rvert^2 \omega_{kl} \mathrm{Re} \bigl[Y_{C}(\omega_{kl}) \bigr],
\end{equation}

with $\Phi_0=h/2e$ the superconducting flux quantum and $\phi$ the reduced (dimensionless) flux operator of the system.
At finite temperature $T > 0$, this is replaced by a relaxation rate 
\begin{equation}
\gamma_{kl}^{\downarrow} = \gamma_{kl} [1 + \bar{n}(\omega_{kl})],
\end{equation}
and an excitation rate
\begin{equation}
\gamma_{kl}^{\uparrow} = \gamma_{kl} \bar{n}(\omega_{kl}),
\end{equation}
with average photon number
\begin{equation}
\bar{n}(\omega) = \frac{1}{e^{\beta \hbar \omega} - 1},
\end{equation}
where $\beta=1/k_b T$. The relaxation times from level $k$ to level $l$ with $k>l$ reported in \cref{tab:par_set} are 

\begin{equation}
T_{1}^{k \mapsto l} = \frac{1}{\gamma_{kl}^{\downarrow}},
\end{equation}

while the excitation times are defined as

\begin{equation}
T_{1}^{l \mapsto k} = \frac{1}{\gamma_{kl}^{\uparrow}}.
\end{equation}

Thus, for any pair of energy levels we take two jump operators $L_{kl}^{\downarrow} = \sqrt{\gamma_{kl}^{\downarrow}} \ket{l}\bra{k}$ and $L_{kl}^{\uparrow} = \sqrt{\gamma_{kl}^{\uparrow}} \ket{k}\bra{l}$, which model relaxation and excitation between the two levels. While for transmon qubits the excitation rate between the first two levels can be neglected at the typical operating temperatures of few $\mathrm{mK}$, this is not the case for the fluxonium. For low frequency fluxonia, the excitation rate $\gamma_{01}^{\uparrow}$ can be comparable to the relaxation rate. Table~\ref{tab:t1_flx} shows some relevant relaxation and excitation times for the fluxonia we considered in this paper.

\section{Details of the microwave pulse and echo}
\label{app:pulse}
In the simulations in the main text we use a piece-wise Gaussian envelope which is defined as 
\begin{widetext}
\begin{equation}
\label{eq:gauss_pulse}
g(t) = \begin{cases}
\frac{1}{1-\exp[-t_{\mathrm{rise}}^2/2 \sigma^2]} \bigl \{\exp[-(t - t_{\mathrm{rise}})^2/2 \sigma^2]-\exp[-t_{\mathrm{rise}}^2/2 \sigma^2] \bigr \}, & 0 \le t < t_{\mathrm{rise}}  \\
1, & t_{\mathrm{rise}} \le t < t_{\mathrm{pulse}} - t_{\mathrm{rise}} \\
\frac{1}{1-\exp[-t_{\mathrm{rise}}^2/2 \sigma^2]} \bigl \{\exp[-(t - (t_{\mathrm{pulse}}- t_{\mathrm{rise}}))^2/2 \sigma^2]-\exp[-t_{\mathrm{rise}}^2/2 \sigma^2] \bigr \}, & t_{\mathrm{pulse}} -  t_{\mathrm{rise}} \le t \le t_{\mathrm{pulse}}, \\
0 & \mathrm{otherwise}.
\end{cases}
\end{equation}
\end{widetext}
using $\sigma = t_\mathrm{rise}/\sqrt{2 \pi}$ and $t_{\mathrm{pulse}}$ the total pulse duration.

In Appendix \ref{app:sw} we have shown that, as in the transmon-transmon case, the CR effect comes with an additional $X$ drive on the transmon. This term commutes with the CR term $X_t Z_f$ and gives rise to an unwanted $X$-rotation on the transmon during the gate. After each simulated gate, we undo this $X$ rotation on the transmon by applying its (noiseless) inverse quantum operation. While an $X$ drive is also present on the fluxonium, the large detuning between the drive frequency and the fluxonium frequency makes the effect of this term negligible.  

In addition, we also consider an echo pulse similar to Refs.~\cite{sheldon2016, malekakhlagh2020,corcoles:RB} with the goal to cancel the $ZZ$ coupling and the $X$-rotation on the transmon qubit during the gate. Let $U_{GP}(t_{\mathrm{pulse}}, \epsd, \omega)$ be the time evolution operator when a Gaussian pulse as in Eq. \eqref{eq:gauss_pulse} with total pulse time $t_{\mathrm{pulse}}$, amplitude $\epsd$ and frequency $\omega$ is applied to the control fluxonium. The echo pulse consists of two pulses with opposite sign of the drive amplitude applied on the fluxonium at chosen frequency $\omega=\omt$ of the target transmon, interleaved with single-qubit $\pi$ rotations around the $X$-axis on the control, fluxonium, qubit $R_X^{(f)}(\pi)$. This gives the time evolution operator

\begin{multline}
U_{\mathrm{echo}} = R_{X}^{(f)}(\pi) U_{GP}(t_{\mathrm{pulse}}, -\epsd, \omt)\times \\
R_{X}^{(f)}(\pi)U_{GP}(t_{\mathrm{pulse}}, \epsd, \omt).
 \end{multline}
In order to implement a CR gate the pulse time of each Gaussian pulse is chosen such that 
\begin{equation}
\label{eq:cr_condition_echo}
\frac{|\mu_{\mathrm{CR}}|}{\hbar} \int_0^{t_{\mathrm{pulse}}} d t g(t) = \frac{\pi}{8},
\end{equation}
so applying essentially half the CR gate. Note that $\epsd \rightarrow -\epsd$ changes the sign $\mu_{CR} \rightarrow -\mu_{CR}$ in Eq.~\eqref{eq:mu_cr_sw} and conjugation by $\pi$-pulses causes $Z_f \rightarrow -Z_f$, so that $U_{\mathrm{echo}}$ implements the CR gate. When we apply $U_{\rm echo}$, both terms $\mu_{X_f} X_f$ and $\mu_{X_t} X_t$ in \eqref{eq:drive-eff-rot} cancel due to $\epsd \rightarrow -\epsd$.

In our simulation we take the single-qubit $\pi$-rotations $R_{X}^{(f)}(\pi)$ to be perfect.
As discussed in Ref.~\cite{malekakhlagh2020}, the echo pulse ideally cancels the effect of the $ZZ$ coupling and the additional $X$-rotations but can also introduce some other unwanted terms in the effective time evolution operator although the overall effect is
positive. In all cases, we always undo the accumulated single-qubit phases via virtual $Z$ gates.

\section{Average number of collisions}
\label{app:num_cols}

\begin{figure}
\vspace{0.3 cm}
\centering
\includegraphics[width=0.4 \textwidth]{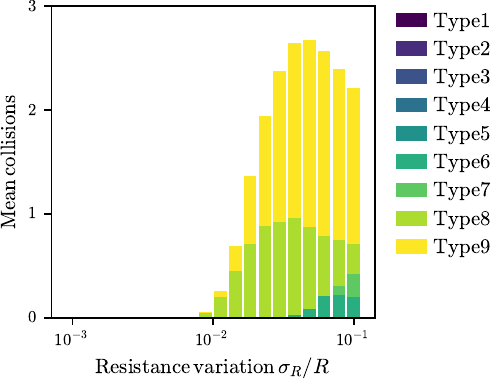}
\caption{Average number of collisions of each type as a function of the resistance variation $\sigma_{R}/R$ for a fixed-frequency transmon-fluxonium $d=3$ surface code lattice and drive amplitudes $\epsd/2\pi = 300\, \mathrm{MHz}$. The number of collisions are collected over $6000$ resamples of the lattice parameters. Collision type 5 is disabled due to the proximity of the transition to that of type 4. Instead the window around collision type 4 accounts for collision 5 as well.}
\label{fig:avg_cols}
\end{figure}

In Sec.~\ref{sec:arch} we obtained the zero-collision device yield based on the frequency collisions and bounds outlined in~\cref{tab:freq_colisions}. In this Appendix we explore the average number of collisions for each type that have occurred for a $d=3$ surface code lattice using a drive amplitude of $\epsd/2\pi = 300\,\mathrm{MHz}$, shown in Fig.~\ref{fig:avg_cols}. We observe that most collisions involve a spectator transmon, specifically, these are collisions of type 8 or 9. Together, these collisions account for most of the reduction in the zero-collision yield observed in Fig.~\ref{fig:freq_cols}. Collision type 8 results in the excitation of a spectator transmon from $\ket{1}$ to $\ket{2}$ during a CR gate. Given the target frequencies in Fig.~\ref{fig:arch_cr}, there exist pairs of transmons ($\omt/2\pi$ and  $\oms/2\pi$ at $4.3\,~\mathrm{GHz}$ and $4.7\,~\mathrm{GHz}$, or $5.3\,~\mathrm{GHz}$ and $5.7\,~\mathrm{GHz}$, respectively) whose frequencies are ideally $100\,\mathrm{MHz}$ away from this collision. Collision type 9 results in the fluxonium qubit leaking from $\ket{0}$ to $\ket{4}$, corresponding to a transition frequency of $\omf{04}/2 \pi = 9.86 \, \mathrm{GHz}$ for the target fluxonium parameters in~\cref{tab:par_set}. In this case, there are pairs of transmon frequencies ($\omt/2\pi$ and  $\oms/2\pi$ at $4.3\,~\mathrm{GHz}$ and $5.7\,~\mathrm{GHz}$, or $4.7\,~\mathrm{GHz}$ and $5.3\,~\mathrm{GHz}$, respectively) the sum of which is $140\,\mathrm{MHz}$ away from this transition. In either case, the variation in the tunnel resistance of $\sigma_{R}/R = 2\%$ translates to a variation in $\omt$ and $\oms$ of about $1\%$ each, which translates to a standard deviation of approximately $50\,\mathrm{MHz}$, leading to the onset of these types of collisions. Collision type 6, which is the next most dominant collision, does not involve any spectator transmons and leads to the $\ket{0}$ to $\ket{5}$ transition on the fluxonium that happens at a frequency $\omf{05}/2 \pi = 13.23 \, \mathrm{GHz}$. For the ideal transmon frequencies, this collision is ideally about $110 \, \mathrm{MHz}$ away from the transmon at frequency $\omt = 4.3 \, \mathrm{GHz}$ and detuned by $300 \, \mathrm{MHz}$ or more from any other transmon. The relatively large number of collisions of type 8 or 9 compared to any other types indicates that the target frequency allocation of the transmons is the main limiting factor behind the current yield.

\bibliography{biblio_flx_transm}

\end{document}